\documentstyle[aps,multicol,epsf,rotate]{revtex}
\begin{document}
\include{psfig}

\newcommand{\en}{\end{equation}}
\newcommand{\ena}{\end{eqnarray}}
\newcommand{\be}{\begin{equation}}
\newcommand{\ee}{\end{equation}}
\newcommand{\bea}{\begin{eqnarray}}
\newcommand{\eea}{\end{eqnarray}}
\newcommand{\bc}{\begin{center}}
\newcommand{\ec}{\end{center}}
\newcommand{\bi}{\begin{itemize}}
\newcommand{\ei}{\end{itemize}}
\newcommand{\il}{$\ell\,\, $}

\draft

\title{Flow Between Two Sites on a Percolation Cluster}

\author{Jos{\'e} S. Andrade Jr.$^{1,2}$, Sergey V. Buldyrev$^1$,
Nikolay V. Dokholyan$^{1,3}$, Shlomo Havlin$^{1,4}$, \\
Peter R. King$^5$, Youngki Lee$^1$,
Gerald Paul$^1$, and H. Eugene Stanley$^1$} 

\address{
$^1$Center for Polymer Studies, Boston University, Boston, MA 02215 \\
$^2$Departamento de F\'{\i}sica, Universidade Federal do Cear\'a,
    60451-970 Fortaleza, Cear\'a, Brazil \\
$^3$Department of Chemistry and Chemical Biology, Harvard University \\
    12 Oxford Street, Cambridge, MA 02138, USA\\
$^4$Minerva Center \& Department of Physics, Bar-Ilan University, Ramat  
    Gan, Israel\\
$^5$Centre for Petroleum Studies, TH Huxley School, Imperial College \\
Prince Consort Road, London SW7 2BP, UK
}

\date{Submitted: 21 April 2000}

\maketitle

\medskip

\begin{abstract}

We study the flow of fluid in porous media in dimensions $d=2$ and
$3$. The medium is modeled by bond percolation on a lattice of $L^d$
sites, while the flow front is modeled by tracer particles driven by a
pressure difference between two fixed sites (``wells'') separated by
Euclidean distance $r$. We investigate the distribution function of
the shortest path connecting the two sites, and propose a scaling {\it
Ansatz\/} that accounts for the dependence of this distribution (i) on
the size of the system, $L$, and (ii) on the bond occupancy
probability, $p$. We confirm by extensive simulations that the {\it
Ansatz\/} holds for $d=2$ and 3, and calculate the relevant scaling
parameters. We also study two dynamical quantities: the minimal
traveling time of a tracer particle between the wells and the length
of the path corresponding to the minimal traveling time ``fastest
path'', which is not identical to the shortest path.  A scaling {\it
Ansatz\/} for these dynamical quantities also includes the effect of
finite system size $L$ and off-critical bond occupation probability
$p$. We find that the scaling form for the distribution functions for
these dynamical quantities for $d=2$ and 3 is similar to that for the
shortest path but with different critical exponents.  The scaling form
is represented as the product of a power law and three exponential
cutoff functions.  We summarize our results in a table which contains
estimates for all parameters which characterize the scaling form for
the shortest path and the minimal traveling time in $2$ and $3$
dimensions; these parameters are the fractal dimension, the power law
exponent, and the constants and exponents that characterize the
exponential cutoff functions.
\end{abstract}

\begin{multicols}{2}

\section{Introduction}

Percolation theory is a paradigmatic model for connectivity, originally
introduced as a mathematical subject in the late 1950s
\cite{Broad57}. Thereafter, percolation theory has been found useful to
characterize many disordered systems
\cite{Stauffer94,Havlin87,Bunde96,Sahimi93,Sahimi94}.

The simplest percolation model is a lattice of bonds occupied with
probability $p$. Neighboring bonds are considered to be {\it
connected\/} if both are occupied.  A set of sites connected by bonds is
called a {\it cluster}. As $p$ increases, new clusters are formed and
previously existing clusters not only grow, but become connected as more
sites are occupied in the system. At a critical value of $p$, $p_c$
(known as the {\it percolation threshold\/}), one spanning cluster
appears and provides overall connectivity. Just at the critical point,
the incipient infinite percolation cluster is an example of a random
fractal that is a useful model for real disordered systems. While the
actual threshold value $p_c$ depends on the particular lattice chosen,
the behavior of the properties measured near the percolation threshold
is universal, depending only on the dimensionality of the system. For
example, the mass $M$ of a cluster diverges at the percolation threshold
as a power $M \sim |p-p_c|^{-\gamma}$ where $\gamma =43/18$ ($d=2$) and
$1.795\pm 0.005$ ($d=3$) regardless of the type of lattice
\cite{Stauffer94,Havlin87,Bunde96}. This universality concept is
extremely useful as it means that we can understand a new system knowing
that its {\it critical exponents\/} are the same as previously-studied
systems. Also, off-lattice continuum percolation appears to have the
same exponent values as those for lattice percolation
\cite{Gawlinski81,Geiger82}.

A comprehensive set of exactly- and numerically-calculated critical
exponents is now available to describe many of the features of
percolation \cite{Stauffer94,Bunde96}, and the percolation paradigm has
been applied to problems of practical interest in heterogeneous
chemistry \cite{Avnir89}, polymer science \cite{PGG92} and transport
phenomena in disordered systems \cite{Ziman79}. In studying transport
phenomena in disordered systems, one must couple additional processes to
the geometrical features of the percolation representation. Typical
examples include the problems of diffusion, reaction and flow through
porous media, as well as the metal-to-insulator transition in polymer
composites and alloys.  These are systems in which the interplay between
structure and phenomenology must be investigated in detail, as it might
be the relevant factor determining optimum material properties.

A useful concept---first put forward by Ambegaokar, Halperin and
Langer \cite{Ambegaokar71}---is that electrical transport in disordered
media with a broad distribution of conductance values is dominated by
those regions where the conductances are larger than some critical value
$\sigma_c$. This critical value is the largest conductance such that the
set of conductances above this threshold value still preserves the
global connectivity of the system.  In percolation terminology, this
cluster would be analogous to the conducting spanning cluster.  It is
then possible to reduce the general problem of transport in a highly
connected disordered media with a broad distribution of conductances to
a percolation problem at criticality. Once the ``critical cluster'' is
identified within the disordered geometry, one can estimate macroscopic
transport properties of the system. This approximation, commonly called
``the critical path method'' \cite{Sahimi94}, has been extended by Katz
and Thompson \cite{Katz86} to estimate transport properties (e.g.,
permeability and electrical conductivity) in disordered materials. Thus,
although the majority of studies of fluid flow through disordered media
are for systems well above the threshold (e.g., sandstone), if the
distribution of permeabilities is sufficiently broad, we can still make
use of percolation concepts to model the relevant geometry of
well-connected disordered media. Therefore, percolation theory is
certainly an appropriate description for a large number of disordered
systems (see also Ref.~\cite{Andrade96} and references therein). This is
the basis of our study and also the basis of the vast literature on
fluid flow in percolation clusters.

The aim of the present paper is to discuss the potential application of
percolation theory as a convenient geometrical model for understanding
numerous aspects of flow through porous rocks
\cite{Dokholyan98c,Lee99}. Special emphasis will be given
to the study of oil displacement, i.e., how hydrocarbons propagate
through geological formations between a pair of wells in the oil
field. This work could also be applied to the breakthrough time for
contamination of a water supply, or the time for released radioactive
material to get from a leaking nuclear repository into the biosphere.

Oil fields are extremely complex, containing geological heterogeneities
on a wide range of length scales from centimeters to kilometers. These
heterogeneities, caused by the sedimentary processes that deposited the
rocks and the subsequent actions on the rock, such as fracturing by
tectonic forces and mineral deposition from aquifer flow, have a
significant impact on hydrocarbon recovery.

The most common method of oil recovery is by displacement. Either water
or a miscible gas (carbon dioxide or methane) is injected in a well (or
wells) to displace the oil to other wells. Ultimately the injected fluid
will break through into a production well where it must be separated
from the oil, which is a very costly process. Once the injected fluid
has broken through, the rate of oil production declines as more injected
fluid is produced. For economic purposes it is important to know when
the injected fluid will break through and what the rate of decline of
oil production will be so that the economic limit of production can be
determined.

Because the sedimentary process that produces the porous rock through
which the fluid flows is very chaotic, the rock is highly
heterogeneous. However, in many cases the rock can be separated into two
types---high permeability (``good'') and low or zero permeability
(``bad'')---and for all practical purposes we can assume the flow takes
place only in the good rock. The spatial distribution of the rock types
is often close to random, in which case the classical percolation
problem is a good approximation. The place of the occupancy probability
$p$ is taken by the volume fraction of the good rock, called the
net-to-gross ratio in the petroleum literature.

We have very little direct knowledge about the spatial distribution of
rock properties in a reservoir. Direct measurements are limited to
samples that represent a fraction of $\approx 10^{-13}$ of the total
reservoir volume.  These samples are taken from the well
locations. Elsewhere the properties have to be inferred from knowledge
of the general geological environment and by analogy with other
reservoirs or surface outcrops.  Hence, there is a great deal of
uncertainty in our prediction of the spatial distribution of these rock
properties. This leads to an uncertainty in our ability to predict the
flow performance, principally the breakthrough time and the production
decline rate. We need to estimate the uncertainty accurately so that
economic risk evaluations can be made.

The conventional approach to this problem is to build a detailed
(numerical) model of the rock properties. These models will honor the
one and two point statistics observed from the wells and analogue
outcrops. The models must also agree with the observed data values at
wells. These models are statistical in nature and conventionally one
samples realizations from the models and performs numerical flow
calculations on the realizations to give a Monte Carlo prediction of
breakthrough and production decline. Unfortunately this process is so
time consuming as to be impractical in many cases. Typically the flow
simulation can take several hours on reasonable workstations. When
hundreds of realizations are sampled to get good statistics, the total
computing time becomes unwieldy. Thus there is a strong need to make
this more efficient so as to come up with very quick, but accurate,
predictions of recovery and the uncertainty due to the lack of knowledge
of the underlying rock properties.

The purpose of this work is to accomplish the goals described above
using methods derived from percolation theory. It is based on two key
assumptions. One is that for many cases the permeability disorder can be
approximated by either permeable or impermeable rock. For example the
reservoir may have been deposited by meandering river belts in which
case the good sand occurs as ``packages'' in an insulating
background. The other assumption is that the flow paths are strongly
controlled by the permeability disorder and not strongly modified by the
flow dynamics themselves. Again there are many cases when this is
reasonable, in particular if the viscosity ratio between the injected
and displaced fluids is not too large or when the system is highly
disordered.

Under these assumptions we can consider the underlying heterogeneity to
be that of a percolating system (not necessarily at threshold).  This
has previously been done to study the fraction of sand connected to well
pairs \cite{King90}. We then look at the dynamic displacement on this
cluster where the flow is controlled by Darcy's law (analogous to Ohm's
law in electrical current flow). We assume that the injected fluid can
be treated as a passive tracer (i.e., one that is not absorbed by the
rock) that is convected along these flow paths. Then the breakthrough
time is the same as the first passage time and is strongly controlled by
the shortest path length between the wells; the post breakthrough
production decline is controlled by the longer paths. Figure~\ref{fig1}a
illustrates a typical percolation backbone and the shortest path between
two points. Note that the lines here represent not microscopic pores but
rather sand bodies whose size are of the order of tens of meters.

The rest of the paper is organized as follows. In Sec.~II, because of
its importance to flow in the cluster, we study the distribution of
shortest paths between two sites in a percolation cluster. In Sec.~III
we study the distribution functions for the dynamic quantities: minimal
traveling time, and length of the path corresponding to minimal
traveling time. Finally, in Sec.~IV, we draw some conclusions and
discuss possible future work.

\section{Shortest path}

This section deals with the distribution of the shortest path between
two sites on a percolation cluster.  Because of the qualitative
resemblance between the shortest path and the minimal traveling time of
a tracer particle, the first step in understanding fluid transport
between two sites in a percolation system is to characterize the
geometrical properties of the shortest connecting path. For example, if
we assume that the traveling time along a path is proportional to the
path length (i.e., all velocities are equal), then we can obtain a rough
estimate for the traveling time from purely geometrical arguments.

\subsection{Basic distribution functions}

The {\it shortest path\/} or {\it chemical distance}, $\ell$, between
two sites on a percolation cluster is defined as the shortest path
connecting the two sites (Fig.~\ref{fig1})
\cite{Alexandrowicz80,Pike81}. The typical value $\ell^*$ of the
shortest path between two sites on a cluster scales with the geometrical
distance, $r$, between these points as
\be
\ell^* \sim r^{d_{\mbox{\scriptsize{\mbox{\scriptsize min}}}}}\, ,
\label{eq:r_l}
\ee
where 
\begin{equation}
\label{e.1x}
d_{\mbox{\scriptsize min}}=\cases{
1.13\pm 0.02 & $[d=2]$\cr
1.374\pm 0.02 & $[d=3]$
}
\end{equation}
is the fractal dimension of the shortest path
\cite{Herrmann88,Grassberger92}.

Consider a hypercubic lattice of $L^d$ sites.  All information about the
distribution of shortest paths is contained in the joint probability
density function $P(r,\ell)$, i.e., the probability that two sites on
the same spanning cluster are separated by geometrical distance $r$ and
chemical path $\ell$. We sum over all chemical paths \il to calculate
the probability distribution that the Euclidean distance between two
sites is $r$,
\be
P(r)\equiv\int P(r,\ell)d\ell \, .
\ee
Similarly, we obtain the probability distribution that two sites are
separated by the chemical distance \il by summing over all possible
geometrical distances,
\be
P(\ell)\equiv\int P(r,\ell)dr \, .
\ee

Given that the shortest distance between these sites is \il, the
conditional probability that the geometrical distance between two sites
is $r$ is \cite{Bunde96}
\be
\label{e4x}
P(r|\ell)=\frac{P(r,\ell)}{P(\ell)} \, .
\ee 
For isotropic media this function has been studied extensively and
$P(r|\ell)$ is of the form \cite{Havlin87,Ray85,Barma86,Neumann88}
\be
P(r|\ell)=A_{\ell} \left(\frac{r}{\ell^{\tilde{\nu}}}\right)^{g_r}
\exp\left( -a
\left(\frac{r}{\ell^{\tilde{\nu}}}\right)^{\tilde{\delta}}\right) \, ,
\label{eq:Prl} \ee
where
\begin{equation}
\label{e6}
\tilde{\delta}={1\over 1-\tilde\nu}={d_{\mbox{\scriptsize min}}\over
d_{\mbox{\scriptsize min}}-1},
\end{equation}
and 
\begin{equation}
\label{e7}
\tilde{\nu}\equiv 1/d_{\mbox{\scriptsize min}}.
\end{equation}
For $d=2$, Ziff recently argued \cite{Ziff99} that
\begin{equation}
\label{e8}
g_r-1=25/24 \qquad\qquad [d=2].
\end{equation}
Our simulations confirm the analytical form of the $P(r|\ell)$ as well
as these values of $\tilde{\nu}$ and $g_r$ (see Fig.~\ref{fig3}a).

The function of interest to us is the conditional probability for
two sites to be separated by the shortest path \il, given that the
geometrical distance between these sites is $r$
\be
\label{e9x}
P(\ell|r)=\frac{P(r,\ell)}{P(r)} \, .
\ee
>From (\ref{e9x}) and (\ref{e4x}), we see that
$P(r|\ell)$ and $P(\ell|r)$ are related
\begin{equation}
\label{e9ax}
P(\ell|r)=P(r|\ell){P(\ell)\over P(r)}.
\end{equation}
At the percolation threshold, it has been shown \cite{Dokholyan98c}
that, in analogy with (\ref{eq:Prl}),
\be
P(\ell|r) \sim \frac{1}{r^{d_{{\mbox{\scriptsize min}}}}}
\left(\frac{\ell}{r^{d_{{\mbox{\scriptsize min}}}}}\right)^{-g_{\ell}}
\exp \left( -a 
\left(\frac{\ell}{r^{d_{{\mbox{\scriptsize min}}}}}\right)^{-\phi_\ell}
\right), 
\label{eq:plr1}
\ee
where
\begin{equation}
\label{e8a}
g_\ell-1={(g_r-1)+(2-d_f)\over d_{\mbox{\scriptsize min}}},
\end{equation}
\be
\phi_\ell=\tilde{\delta}\tilde{\nu}=\tilde{\nu}/(1-\tilde{\nu})={1\over 
d_{{\mbox{\scriptsize min}}}-1},
\label{eq:phi_l}
\ee
and
\begin{equation}
\label{e10x}
d_f=\cases{
91/48 & $[d=2]$\cr
2.524\pm 0.008 & $[d=3]$
}
\end{equation}
is the fractal dimension of the incipient infinite cluster
\cite{Bunde96}.  Substituting (\ref{e8}) into (\ref{e8a}), we find for
$d=2$
\begin{equation}
\label{e13a}
g_\ell=2.01\pm 0.02 \qquad\qquad [d=2].
\end{equation}

The probability distribution of more practical interest is $P'(\ell|r)$,
defined in the same way as $P(\ell|r)$ but for any two randomly-chosen
points separated by geometrical distance $r$ and on the same cluster,
but not necessarily on the incipient infinite cluster
\cite{Dokholyan98c}. As shown in Appendix~\ref{appendix_C}, $P'(\ell|r)$
has the same scaling form as in Eq.~(\ref{eq:plr1}), but with $g_\ell$
replaced by
\be
g_{\ell}'=g_{\ell}+{d-d_f\over d_{\mbox{\scriptsize min}}}.
\ee
Figure~\ref{fig3}b illustrates the difference between $g_\ell$ and
$g'_\ell$.

\subsection{Distribution of shortest path}

The complete scaling form of $P'(\ell|r)$, which accounts also for
finite size effects and off-critical behavior, has been studied for
$d=2$ and reported in \cite{Dokholyan98c}. Specifically, the following
{\it Ansatz\/} has been proposed \cite{Dokholyan98c}
\begin{eqnarray}
\nonumber P'(\ell|r) &\sim& {1 \over r^{d_{\mbox{\scriptsize min}}}}\left({\ell
\over r^{d_{\mbox{\scriptsize min}}}}\right)^{ -g_{\ell}'}
f_1\left({\ell \over r^{d_{\mbox{\scriptsize min}}}}\right)\cdot\\
&~& f_2\left({\ell \over L^{d_{\mbox{\scriptsize min}}}}\right)~
f_3\left({\ell \over \xi^{d_{\mbox{\scriptsize min}}}}\right)\, ,
\label{eq:minansatz}
\end{eqnarray}
where $\xi \sim |p-p_c|^{-\nu}$ is the pair connectedness length, and
the scaling functions have the form
\begin{equation}
\label{e.12a}
f_1(x)\equiv\exp(-ax^{-\phi}),
\end{equation}
\begin{equation}
\label{e.12b}
f_2(x)\equiv\exp(-bx^{\psi}),
\end{equation}
and
\begin{equation}
\label{e.12c}
f_3(x)\equiv\exp(-cx). 
\end{equation}

The function $f_1$ accounts for the lower cut-off due to the constraint
$\ell>r$, while $f_2$ and $f_3$ account for the upper cut-offs due to
the finite size effect and due to the finite correlation length
respectively. Either $f_2$ or $f_3$ becomes irrelevant, depending on the
magnitudes of $L$ and $\xi$: for $L<\xi$, $f_2$ dominates the upper
cut-off, otherwise $f_3$ dominates. We assume the independence of the
finite size effect and the effect of the concentration of the occupied
sites, so that Eq.~(\ref{eq:minansatz}) can be represented as a product
of the terms which are responsible for the finite size effect ($f_2$)
and the effect of the concentration ($f_3$). Simulations for $d=2$
\cite{Dokholyan98c} have been used to test this assumption.

\subsection{Shortest path in three dimensions}

\subsubsection{Behavior at criticality}

Here we extend the study of $P'(\ell|r)$ to $d=3$. We numerically test
the scaling conjecture (\ref{eq:minansatz}) exactly at the percolation
threshold $p=p_c$---in which case $\xi=\infty$ so
$f_3=f(0)=\mbox{const}$. We build clusters using the Leath algorithm
\cite{Alexandrowicz80,Pike81,Leath76}.  Since the Leath algorithm
corresponds to the process of selecting a random point on the lattice,
the probability $P'(\ell|r)$ is equal to the probability that a pair of
randomly selected points has chemical distance $\ell$ and geometrical
distance $r$, given that they belong to the same cluster, a cluster that
is not necessarily the infinite cluster. Hence Eq.~(\ref{eq:minansatz})
reduces to
\begin{eqnarray}
\nonumber P'(\ell |r) &\sim& {1 \over r^{d_{\mbox{\scriptsize min}}}}\left({\ell \over
r^{d_{\mbox{\scriptsize min}}}}\right)^{ -g_{\ell }'}\cdot\\
&~& f_1\left({\ell \over r^{d_{\mbox{\scriptsize min}}}}\right)~
f_2\left({\ell \over L^{d_{\mbox{\scriptsize min}}}}\right) \quad
[p=p_c].
\label{eq:minansatz1}
\end{eqnarray}

Figure~\ref{fig5}a shows that, in the range $r^{d_{\mbox{\scriptsize
min}}}<\ell <L^{d_{\mbox{\scriptsize min}}}$, $P'(\ell |r)$ has
power-law behavior with slope
\begin{equation}
\label{e15a}
g_{\ell }'=2.3 \pm 0.1, \qquad\qquad [d=3]
\end{equation}
and rapidly vanishes for $\ell<r^{d_{\mbox{\scriptsize min}}}$ and for
$\ell>L^{d_{\mbox{\scriptsize min}}}$. To determine the functions $f_1$
and $f_2$, we compute the rescaled probability distribution
\be 
\Phi\left({\ell\over r^{d_{\mbox{\scriptsize min}}}}\right) \equiv
P'(\ell |r)~\ell^{g_\ell'}~r^{-d_{\mbox{\scriptsize min}}(g_\ell-1)},
\label{eq:rescaled} 
\ee
and plot it against scaling variable
$x\equiv\ell/r^{d_{\mbox{\scriptsize min}}}$ (see Fig.~\ref{fig5}b)
using the value $d_{\mbox{\scriptsize min}}=1.374$. According to
Eq.~(\ref{eq:minansatz1})
\be
\Phi(x)=Af_1(x)f_2\left[ x\left({r\over L}\right) ^{d_{\mbox{\scriptsize
min}}}\right].  
\label{eq:Phix}
\ee
Therefore, $\Phi(x)$ should depend only on $x$ and the ratio $r/L$.
Indeed, Fig.~\ref{fig5}b shows excellent data collapse for $L/r=8$,
with sharp cutoffs governed for $x<1$ by $f_1(x)$ and for
$x>(L/r)^{d_{\mbox{\scriptsize min}}}$ by
$f_2[x(r/L)^{d_{\mbox{\scriptsize min}}}]$.

In order to test the assumption that the functions $f_1$ and $f_2$ are
stretched exponentials with exponents $\phi_{\ell }$ and $\psi_{\ell }$,
we plot 
\begin{equation}
\label{e17a}
\Pi(x) \equiv \log_{10}[A/\Phi(x)]
\end{equation}
versus $x$ in double logarithmic scale for various values of
normalization constant $A$ (see Fig.~\ref{fig5}c). If the stretched
exponential conjecture is correct, $\Pi(x)$ should have two straight
line asymptotes for $\log x \rightarrow +\infty$ with the slope
$\psi_{\ell }$ and for $\log x \rightarrow -\infty$ with the slope
$-\phi_{\ell }$. We find that the slopes $\phi_{\ell }$ and
$\psi_{t_{\mbox{\scriptsize min}}}$ of the straight line fits depend
weakly on the value of $A$.  Using $A=0.08$, we obtain the longest
regimes of straight line behavior. For this value of $A$, we find
$\phi_{\ell } \approx 2.1$ and $\psi_{\ell }\approx 2.5$.  Equation
(\ref{eq:phi_l}) yields a predicted value of $\phi_\ell =2.67$ in good
agreement with our simulation result.

\subsubsection{Off-critical behavior}

For $p\neq p_c$, we identify three regimes determined by the value of
the connectedness length, $\xi$, in relation to the values of $r$ and $L$:

\begin{itemize}

\item[{(i)}] $\xi>L>r$. In this regime, the fact that $p\neq p_c$ cannot
be detected because the connectedness length is larger than the other
relevant variables.

\item[{(ii)}] $L>\xi>r$.  In this case, the upper cutoff of the distribution
Eq.~(\ref{eq:minansatz}) is governed by $f_3$ and the functional form of
the rescaled probability $\Phi$ is given by
\be
\Phi(\ell/r^{d_{\mbox{\scriptsize min}} })
\sim f_1\left({\ell\over r^{d_{\mbox{\scriptsize min}} }}\right)
f_3\left({\ell\over\xi^{d_{\mbox{\scriptsize min}} }}\right).
\label{eq:phix}
\en    
For large $\ell$, we suggest an exponential decay of $\Phi$
\be
\Phi(\ell/r^{d_{\mbox{\scriptsize min}} })
\sim \exp\left(-c{\ell\over\xi^{d_{\mbox{\scriptsize min}}}}\right). 
\label{eq:plrp}
\en
Indeed, for $p<p_c$, semi-logarithmic plots of $\log
\Phi(\ell/r^{d_{\mbox{\scriptsize min}} })$ versus $\ell$ shown in
Fig.~\ref{fig6}a can be approximated by straight lines with slopes which
approach zero as $p \rightarrow p_c$.  According to Eq.~(\ref{eq:plrp}),
these slopes $k(p)$ should be proportional to
$\xi^{-d_{\mbox{\scriptsize min}}}\sim |p-p_c|^{d_{\mbox{\scriptsize
min}}\nu}\approx |p-p_c|^{1.19}$. Figure~\ref{fig6}b shows a double
logarithmic plot of $|k(p)|$ versus $|p-p_c|$ for $p<p_c$.  This curve
can be well approximated by a straight line with slope 1.22 in good
agreement with the scaling conjecture (\ref{eq:minansatz1}).  For
$p>p_c$ a similar analysis should hold. However, limitations on the size
of the system we can simulate make the analysis
problematic. Figure~\ref{fig6}c shows $P'(\ell)$ for various values of
$p>p_c$. Note that it is only for values of $p\geq p_c+0.03$ that the
distributions ``cut-off'' at smaller $\ell$ than the distribution for
$p=p_c$. Thus it is only for values of $p-p_c\geq 0.03$ that the large
$\ell$ behavior of Eq.~(\ref{eq:minansatz}) is determined by the fact
that the system is not at criticality (i.e., by $f_3$) as opposed to
being determined by the finite size of the system (i.e., by $f_2$).
Below $p=p_c+0.03$, $\xi$ is still greater than $L$. On the other hand,
if $p$ is not close to $p_c$, the scaling form is not expected to
hold. Thus, the results are inconclusive based on the sizes of the
systems we can generate---we cannot determine the parameters that govern
the large $\ell$ behavior of Eq.~(\ref{eq:minansatz}) above $p_c$.

\item[{(iii)}] $L>r>\xi$. When the connectedness length $\xi$ is
smaller than the distance $r$ between the wells, the system can be
considered homogeneous \cite{Koplik94,King99}. This can be seen in
Fig.~\ref{fig7}a in which we plot $P(\ell|r)$ for various values of
$r$ at $p=0.7$ for 2d site percolation ($p_c=0.593$).  As $r$
increases from below to above the connectedness length, the form of
the distribution changes from the power law distribution of
Eq.~(\ref{eq:minansatz}) to a Gaussian distribution with a pronounced
peak, a characteristic of homogeneous systems. Furthermore, as shown
in Fig.~\ref{fig7}b, the fractal dimension of the shortest length
crosses over from $d_{\mbox{\scriptsize min}}=1.13$ to
$d_{\mbox{\scriptsize min}}=1.0$, characteristic of a homogeneous
system \cite{Koplik94,King99}. The convergence to a Gaussian can be
expected due to the following considerations. The minimal path
connecting the wells separated by distance $r$ passes through $r/\xi$
independent blobs.  For each of these blobs, the probability
distribution for the shortest path across the blob, $\ell_b$, is still
given by Eq.~(\ref{eq:minansatz}), but with $r$ and $L$ replaced by
$\xi$ and $\ell$ replaced by $\ell_b$. This distribution is
characterized by $\langle\ell_b\rangle\sim\xi^{d_{\mbox{\scriptsize
min}}}$ and variance
$\sigma_b^2\equiv\langle\ell_b^2\rangle-\langle\ell_b\rangle^2\sim
\xi^{2d_{\mbox{\scriptsize min}}}$. The total minimal path is the sum
of $n=r/\xi$ independent variables $\ell_b$, hence it converges to a
Gaussian with
\begin{equation}
\label{e26x}
\langle\ell\rangle\sim r\xi^{d_{\mbox{\scriptsize
min}}-1}\qquad\mbox{and}\qquad\sigma^2\sim r\xi^{2d_{\mbox{\scriptsize
min}}-1}.
\end{equation}
Thus the slope of the graph, $k(p)$, of $\langle\ell\rangle$ vs $r$ in Fig.~5c
should decay as 
\begin{equation}
\label{e27x}
k(p) \sim |p-p_c|^{-\nu(d_{\mbox{\scriptsize min}}-1)}=|p-p_c|^{-0.17}
\end{equation}
and the slope of $\sigma^2$ versus $r$ should decay as
\be
|p-p_c|^{-\nu(2d_{\mbox{\scriptsize min}}-1)}=|p-p_c|^{-1.7}.
\label{e27y}
\en
Indeed, (see Fig~\ref{fig7}d) we see that the slope of $\sigma^2$
versus $r$ decays with $p$ more strongly than that of
$\langle\ell\rangle$ versus $r$. The numerical values of slopes from
Figs.~\ref{fig7}c and ~\ref{fig7}d are in good agreement with the
theoretical predictions Eqs.~(\ref{e27x},\ref{e27y}). For $d=3$ we
expect similar behavior.

\end{itemize}

\subsection{Rectangular boundary conditions}

Since realistic oil fields do not  have square boundaries, it is reasonable
to ask what is the effect of rectangular boundary conditions. Here we study the
distributions of the shortest length and minimal time on a
two-dimensional lattice of size $L_x \times L_y$, $L_x \ne L_y$. We
position the wells at points $A$ and $B$ separated by distance $r=16$
along $x$-axis. We study two cases (Fig.~\ref{fig8}):

\bi
\item[{(a)}] $L_x$ is fixed ($L_x=32$), and we vary $L_y$ ($L_y=64$,
128, 256, 512, 1024); 
\item[{(b)}] $L_y$ is fixed ($L_y=32$), and we vary $L_x$ ($L_x=64$,
128, 256, 512, 1024).
\ei

We find that (i) the shortest length and minimal time distributions are
identical in all of the above cases and (ii) the scaling form for these
anisotropic cases is the same as the isotropic case with $L$ replaced by
the minimum of $L_x$ and $L_y$, i.~e. the scaling form of the
distribution Eq.~(\ref{eq:minansatz}) remains unchanged with the exception
that $L$ is determined by
\be
\label{e20x}
L=\min (L_x, L_y)\, .
\ee

Equation (\ref{e20x}) can be a result of competing exponentials $f_{2x}$
and $f_{2y}$. Since both $f_{2x}$
and $f_{2y}$ are rapidly decaying functions with $L_x$ and $L_y$ [see
Eq.~(\ref{e.12b})], the finite-size cutoff of $P(\ell|r)$ is determined
by the {\it smaller\/} of $L_x$ and $L_y$.
The fact that the results are independent of the axis along which the
wells are aligned can be understood by realizing that the probability
that ``oblong'' clusters connect two points separated by a distance
longer than the minimum of $L_x$ and $L_y$ is low. This finding has the
implication that in anisotropic fields, a number of well pairs would be
needed to optimize the recovery of oil in the field.

\section{Minimal Traveling Time and Fastest Path}

We turn next to dynamics, the study of flow on percolation clusters,
which has close ties to such applications as hydrocarbon recovery and
ground-water pollution
\cite{Bunde96,Saffman59,Koplik88,Bacri90,Sahimi95}. In this section, we
discuss the properties of the flow on $d=2$ and $d=3$ bond percolation
clusters. Specifically, we investigate the scaling properties of the
distributions of {\it minimal traveling time\/} and {\it the length of
the path corresponding to the minimal traveling time\/} (fastest path)
of the tracer particles. Some of the results in $d=2$ were reported
previously \cite{Lee99}.  Here we extend the work to $d=3$, and study
the effects of finite system size and off-criticality for $d=2$ and
$d=3$.

\subsection{The model}

We study incompressible flow between two sites $A$ and $B$ separated by
Euclidean distance $r$ (see Fig.~1). To model the flow front, we use
passive tracers---particles not absorbed by the surroundings, that move
only by convection, ignoring molecular diffusion which is slow on the
time scales of interest.  The convection is governed by the flow field
due to the pressure difference between sites connected by the bonds. We
simulate the flow of a tracer particle starting from the injection point
$A$ traveling through the medium along a path connected to the recovery
point $B$.  The dynamics of flow at a macroscopic level on the
percolation cluster is determined by the local flow (local currents) on
the individual bonds in the backbone of the cluster.  The velocity of a
tracer at each bond is determined by the pressure difference across that
bond (Darcy's law \cite{Dullien79}):
\be
v_{ij} = T (P_j - P_i)\, ,\label{eq:vij-P}
\ee
where $P_i$ and $P_j$ are the values of pressure at sites $i$ and $j$.
The coefficient $T$, which is a function of permeability $k$, viscosity
$\eta$, and the length of a bond $L_b$ ($T=k/(\eta L_b)$), is set to
1. We normalize the velocities assuming the total flow $J$
between $A$ and $B$ is fixed, independent of the distance between $A$
and $B$, and the realization of the porous media. This resembles more
closely oil recovery processes where constant flow, as opposed to
constant pressure, is maintained.

\begin{figure}[htb]
\centerline{
\epsfxsize=15.0cm
\epsfbox{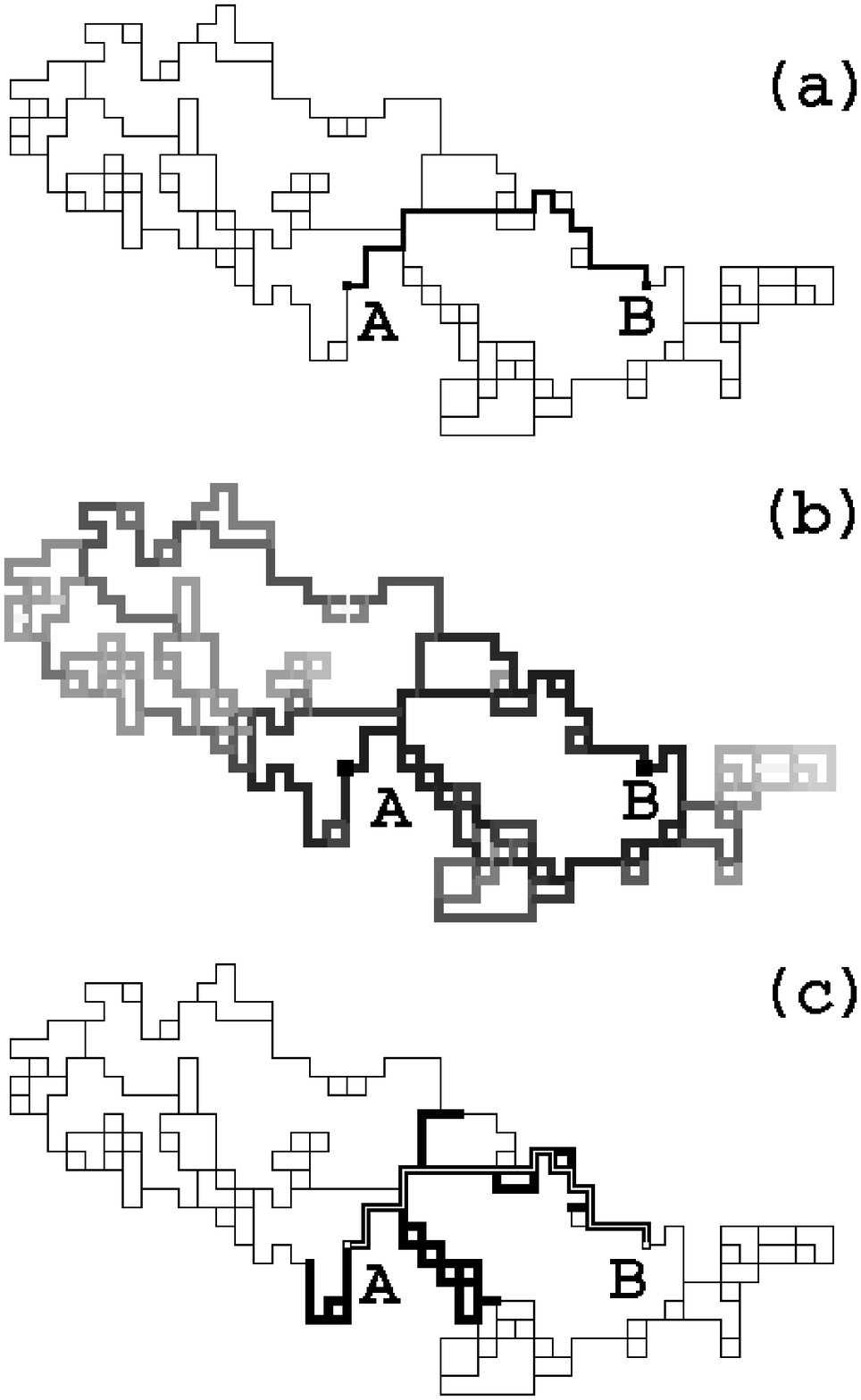}
\vspace*{1.0cm}
}
\caption
{
%
%
%
An example of (a) percolation backbone (thin lines) and shortest path
(thick line) corresponding to the flow between points $A$ and $B$---note
that the lines here represent not microscopic pores but rather sand
bodies whose size are of the order of tens of meters; (b) magnitudes of
currents on cluster backbone are depicted in gray scale with the
lightest areas corresponding to the smallest currents and the darkest to
the largest currents; and (c) time evolution of the flow between points
$A$ and $B$ on a two-dimensional percolation cluster. Thick lines denote
bonds reached by tracer particles at time $t_{\mbox{\scriptsize min}}$.
The double line between $A$ and $B$ denotes the ``fastest path'' between
these two points.
}
\label{fig1}
\end{figure}

\begin{figure}[htb]
\centerline{
\epsfxsize=8.0cm
\rotate[r]{\epsfbox{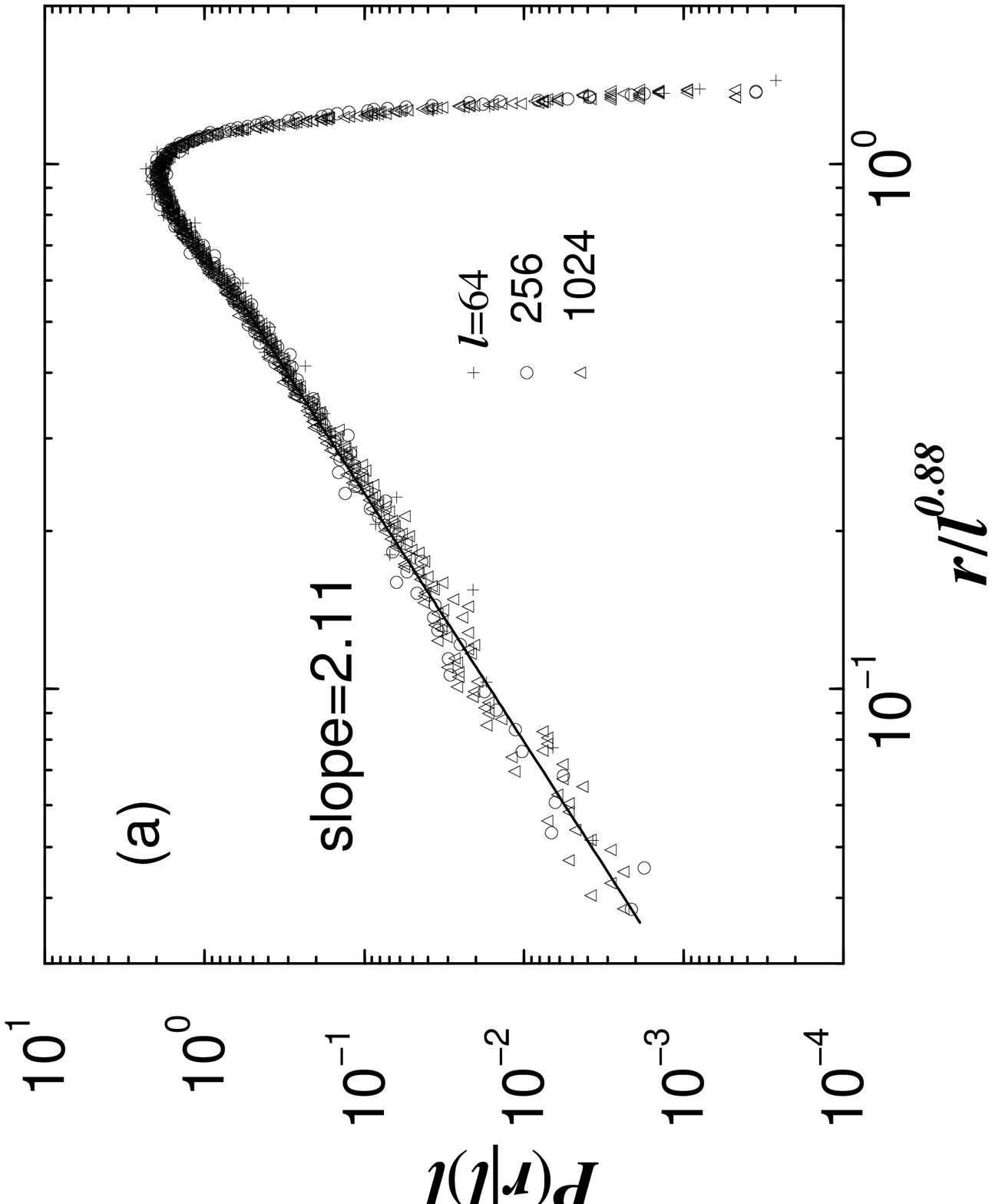}}
\vspace*{1.0cm}
}
\centerline{
\epsfxsize=8.0cm
\rotate[r]{\epsfbox{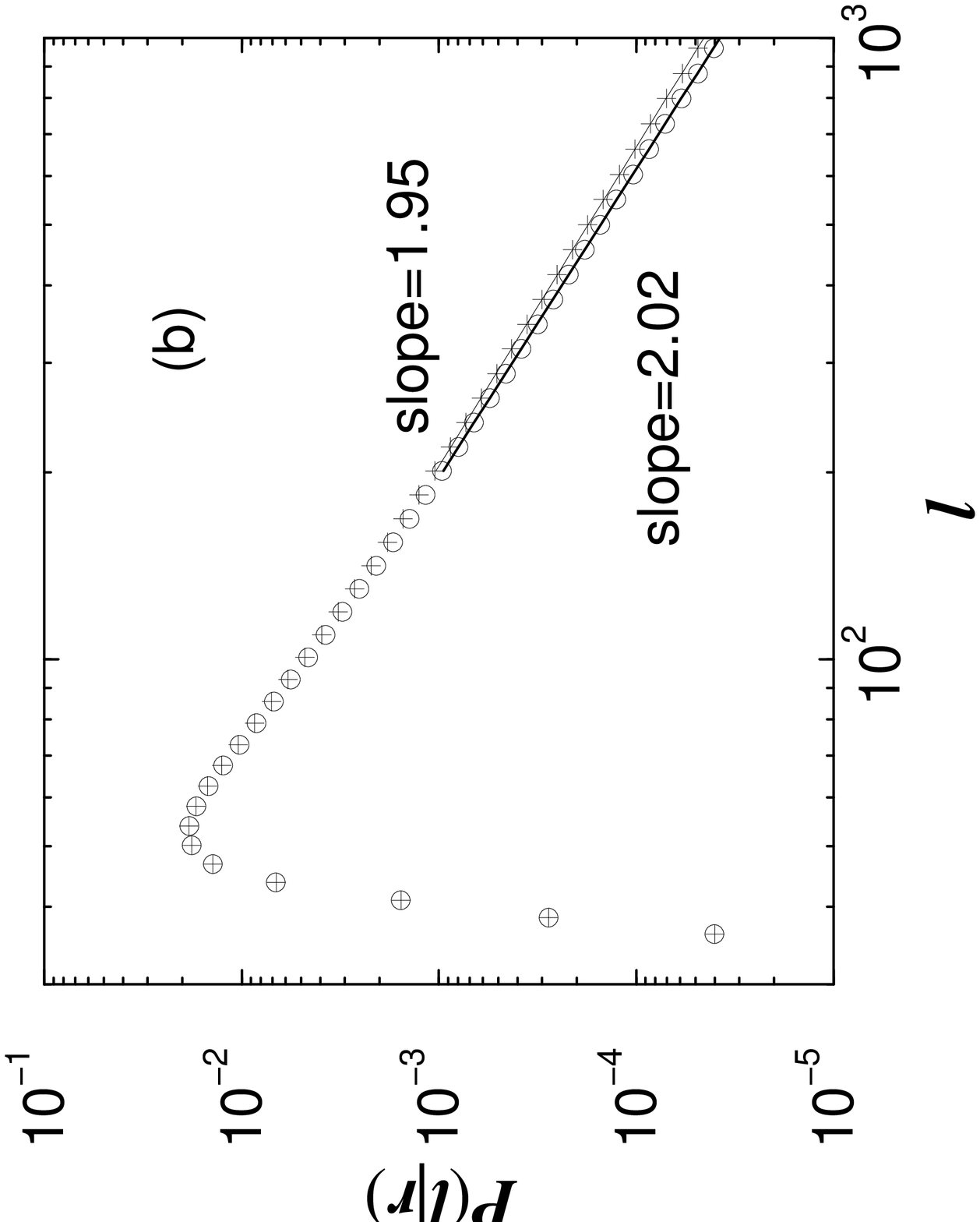}}
\vspace*{1.0cm} }
\caption{
(a) Log-log plot of $\ell^{\nu}P(r|\ell)$ for two-dimensional
percolation at the percolation threshold for $\ell=64$ ($\bigcirc$), 256
($\Box$) and 1024 ($\bigtriangleup$). The best data collapse is obtained
with $\nu=0.88$ and the slope of the power regime is $2.11$. (b) Log-log
plot of $P(\ell|r)$ ($+$) and $P'(\ell|r)$ ($\bigcirc$) for
two-dimensional percolation at criticality and for the system size
$L=1024$ and the distance between wells is $r=32$. The power-law regime
of $P'(\ell|r)$ has slope $g_{\ell}'=2.02$ (thick solid line), while
that of $P(\ell|r)$ has slope $g_{\ell}=1.95$ (thin solid line).
The purpose of this figure is to illustrate the difference between
$g_\ell$ and $g_\ell'$. The values shown here are lower than those
predicted by Ziff \protect\cite{Ziff99} because we have not included
corrections to scaling in our determination of these quantities.
}
\label{fig3}
\end{figure}

\begin{figure}[htb]
\centerline{
\epsfxsize=7.0cm
\rotate[r]{\epsfbox{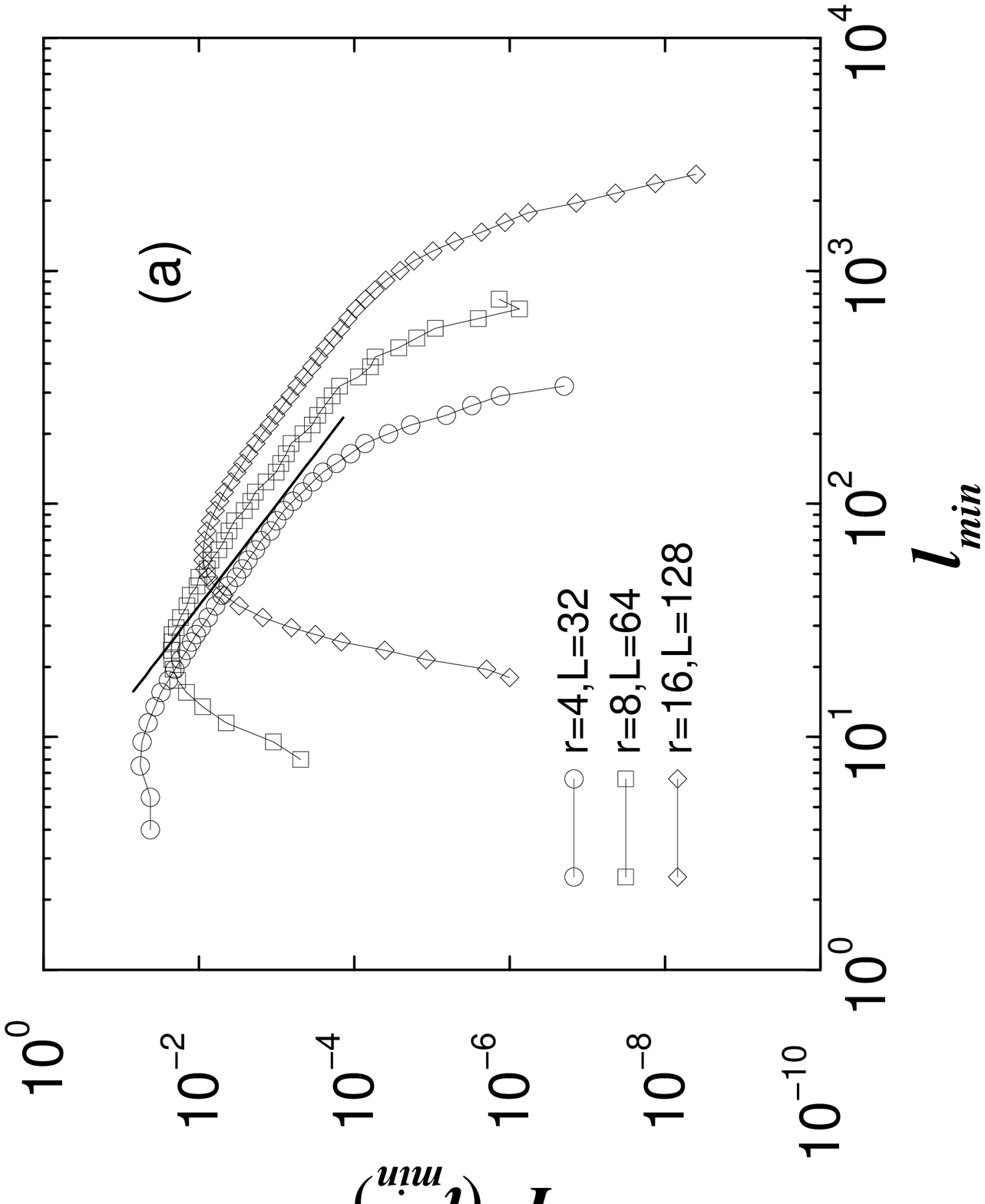}}
}
\centerline{
\epsfxsize=7.0cm
\rotate[r]{\epsfbox{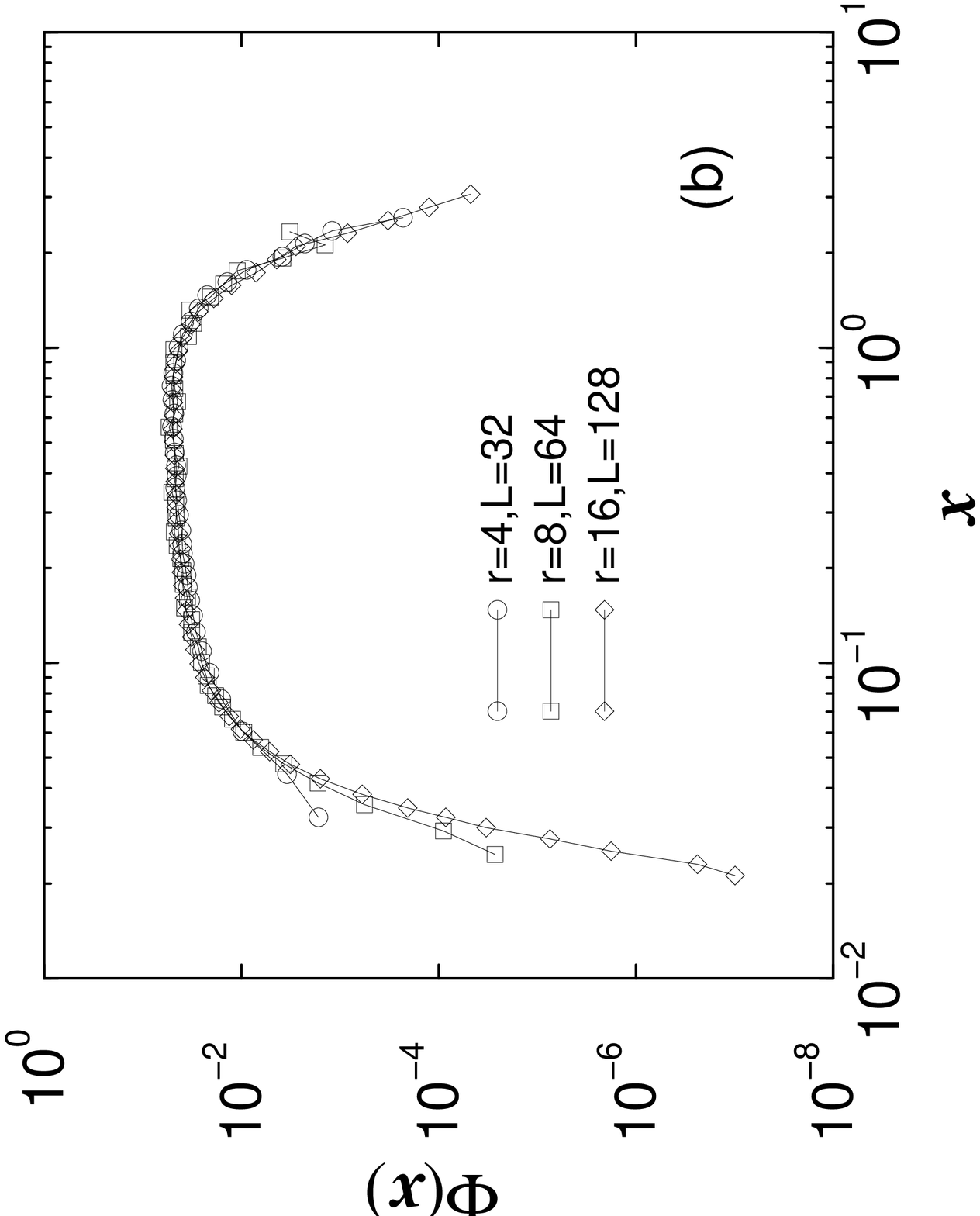}}
}
\centerline{
\epsfxsize=7.0cm
\rotate[r]{\epsfbox{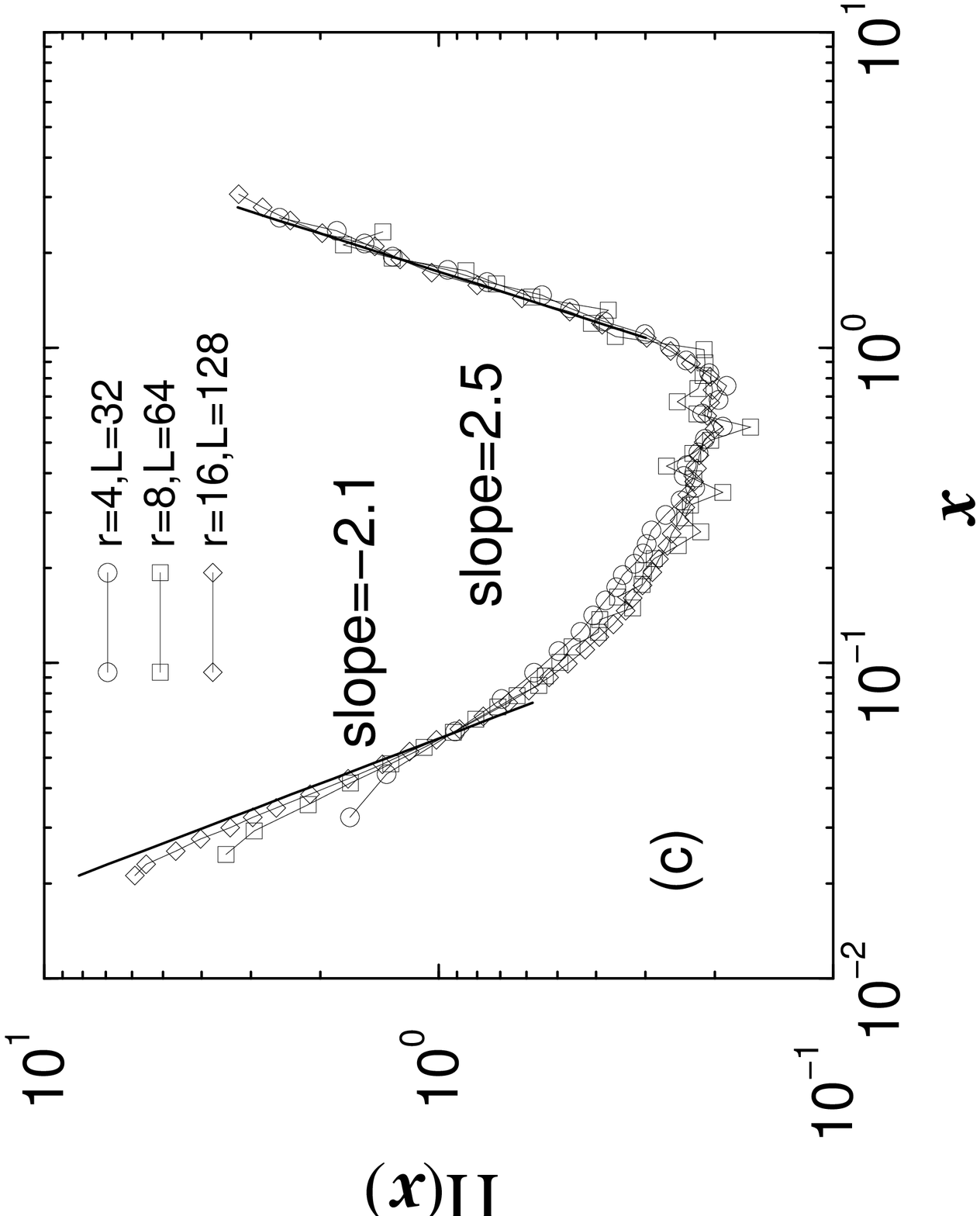}}
\vspace{3.0cm}
}
\caption{For $d=3$, (a) log-log plot of $P'(\ell|r)$ at criticality
($p=p_c\approx 0.2488$) and for different sets of parameters:
$(r,L)=(2,32),(4,64),(8,128)$. The straight line regime has slope
$g_{\ell}'=2.3$.  (b) Log-log plot of rescaled probability
$\Phi(x)\equiv P'(\ell |r) ~ x^{g_{\ell}'} ~ r^{d_{\mbox{\scriptsize
min}}}$ against rescaled length $x\equiv\ell/r^{d_{\mbox{\scriptsize
min}}}$ using the values, $g_{\ell}'=2.3$ and $d_{\mbox{\scriptsize
min}}=1.39$. The curves are flat in the center because $f_2(x)$ is a
stretched exponential (see Eq.~({\protect{\ref{eq:Phix}}})). (c)
Log-log plot of transformed probability $\Pi(x)=\log_{10}[A/\Phi(x)]$
versus $x=\ell/r^{d_{\mbox{\scriptsize min}}}$. The slopes of the
solid lines give the power of the stretched exponential function $f_1$
and $f_2$ in Eq.~({\protect{\ref{eq:Phix}}}). Using the parameter
$A=0.08$, the slopes give $\phi \approx 2.1$ for the lower cut-off and
$\psi\approx 2.5$ for the upper cut-off.}
\label{fig5}
\end{figure}

\begin{figure}[htb]
\centerline{
\epsfxsize=7.0cm
\rotate[r]{\epsfbox{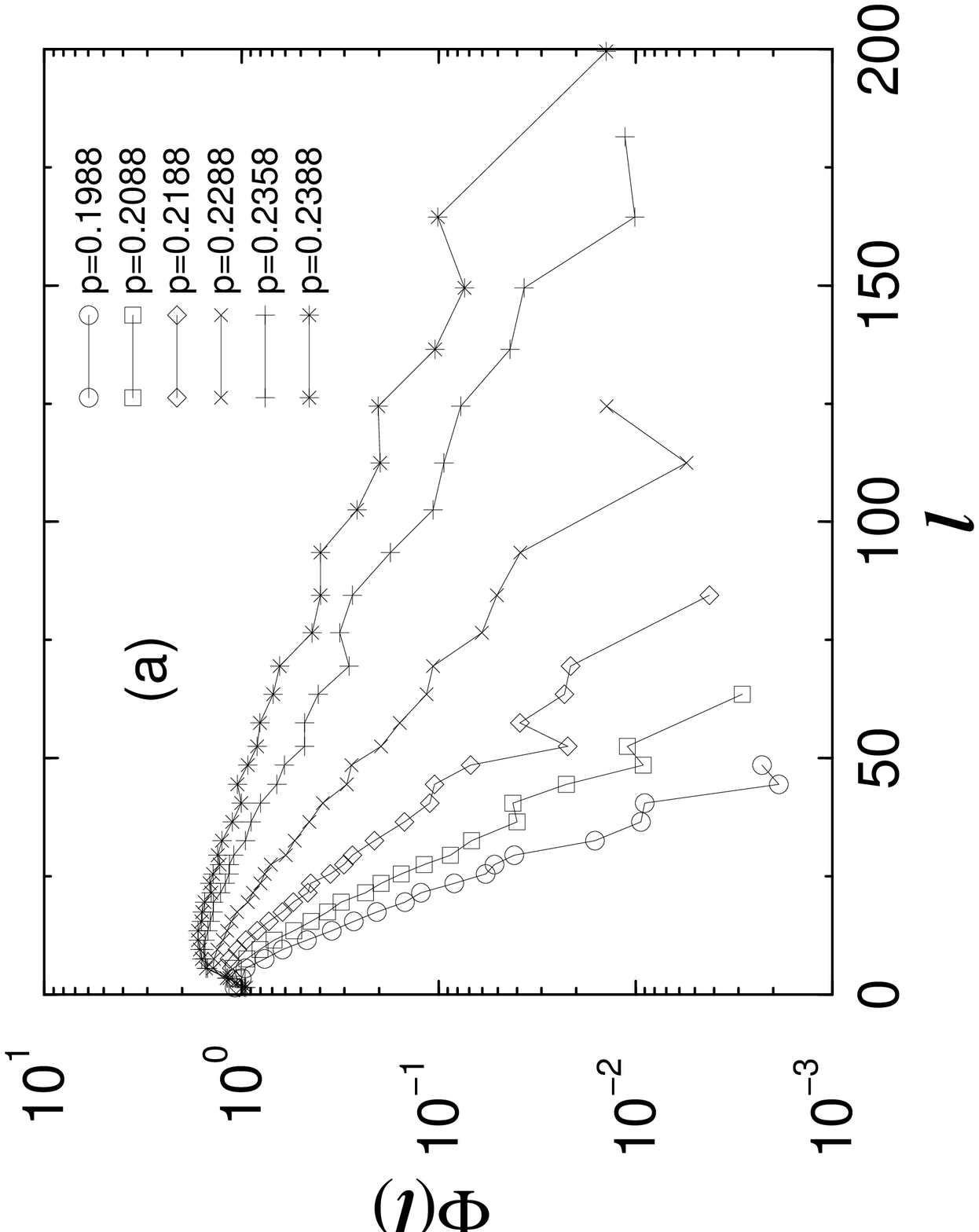}}
}
\centerline{
\epsfxsize=7.0cm
\rotate[r]{\epsfbox{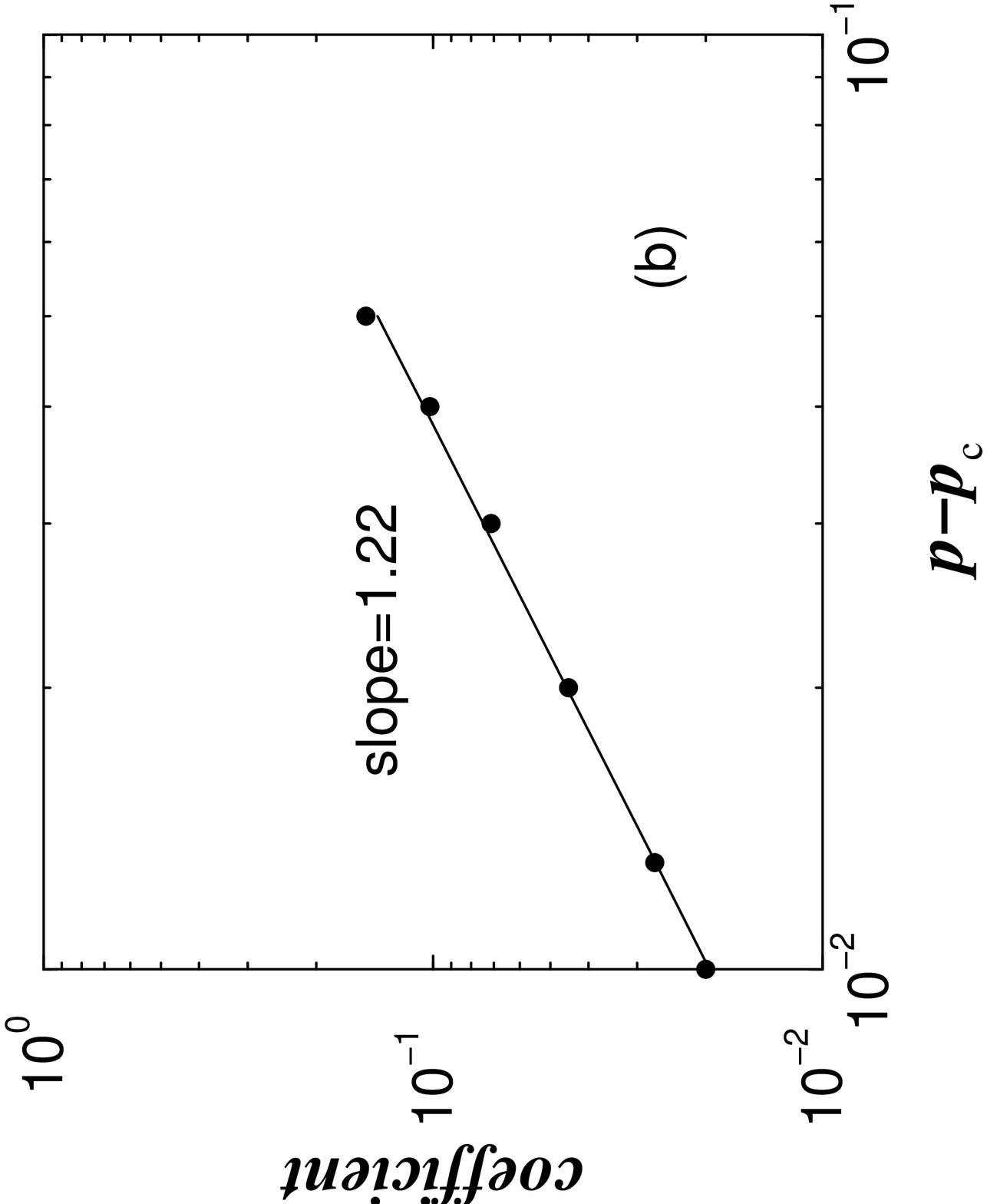}}
}
\centerline{
\epsfxsize=7.0cm
\rotate[r]{\epsfbox{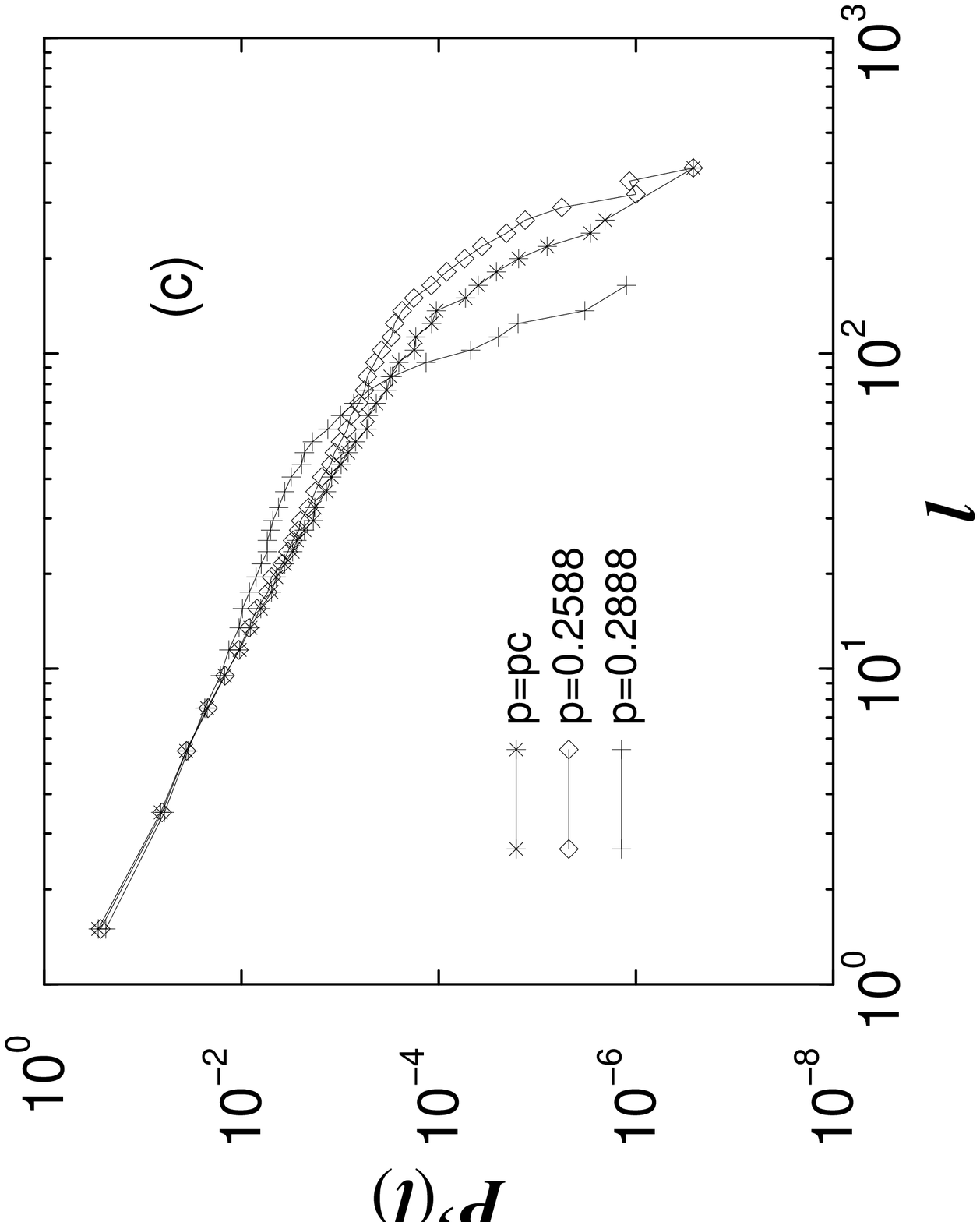}}
}
\vspace{1.0cm}
\caption{
For $d=3$, (a) semi-logarithmic plot of transformed probability
$\Phi(\ell)$ (see Eq.~(\protect{\ref{eq:Phix}})) versus $\ell$ shows
pure exponential behavior of $f_3$. (b) The slope of the log-log plot of
the coefficient in exponential function $f_3$ as a function of $|p-p_c|$
gives the value $\nu d_{\mbox{\scriptsize min}} \approx 1.22$ for
$p<p_c$. (c) $P'(\ell)$ for $p>p_c$. Note that it is only for $p\geq
0.2788$ that the large $\ell$ behavior is determined by the fact that
the system is not at criticality.
}
\label{fig6}
\end{figure}

\begin{figure}[htb]
\centerline{
\epsfxsize=8.0cm
\rotate[r]{\epsfbox{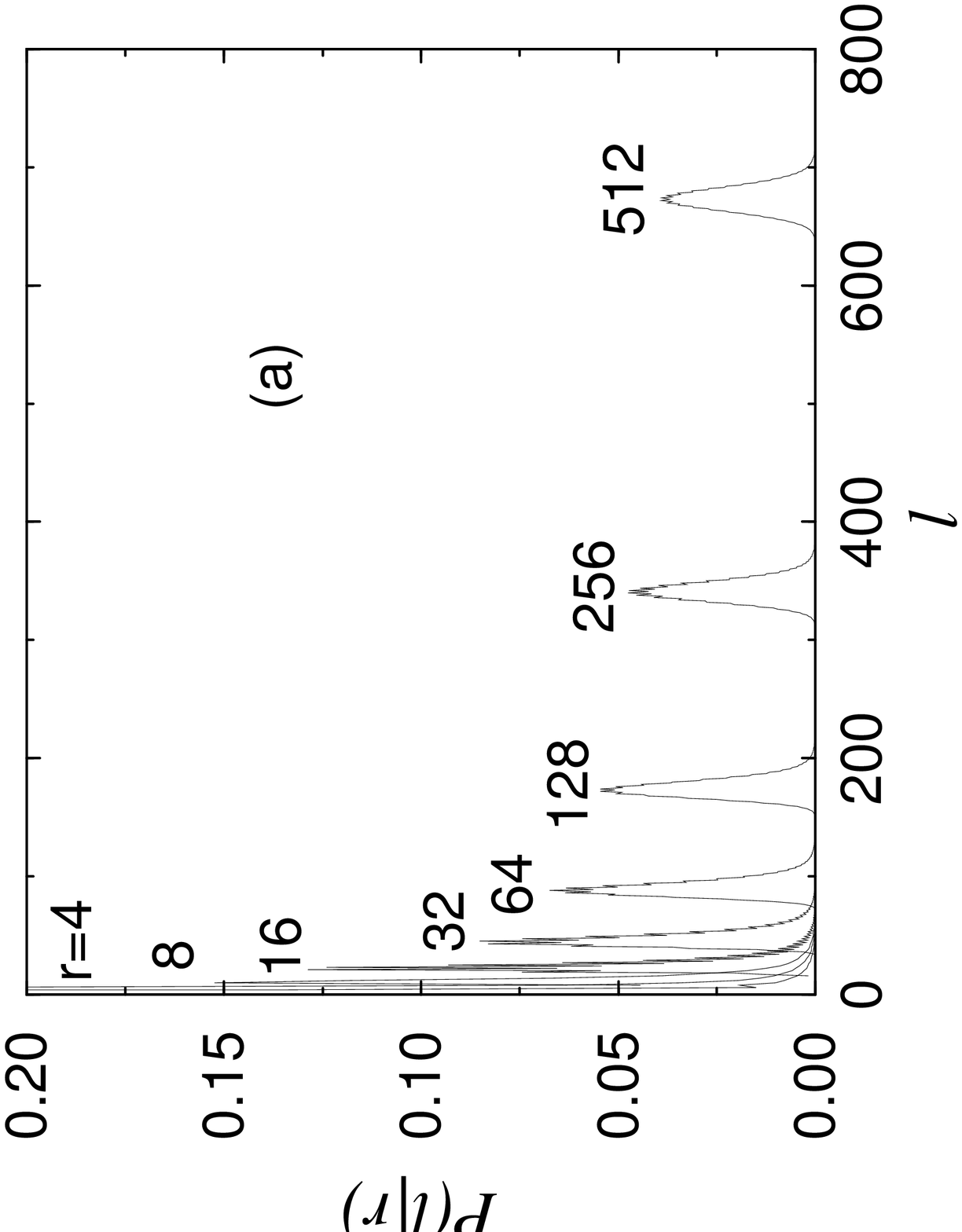}}
}
\centerline{
\epsfxsize=8.0cm
\rotate[r]{\epsfbox{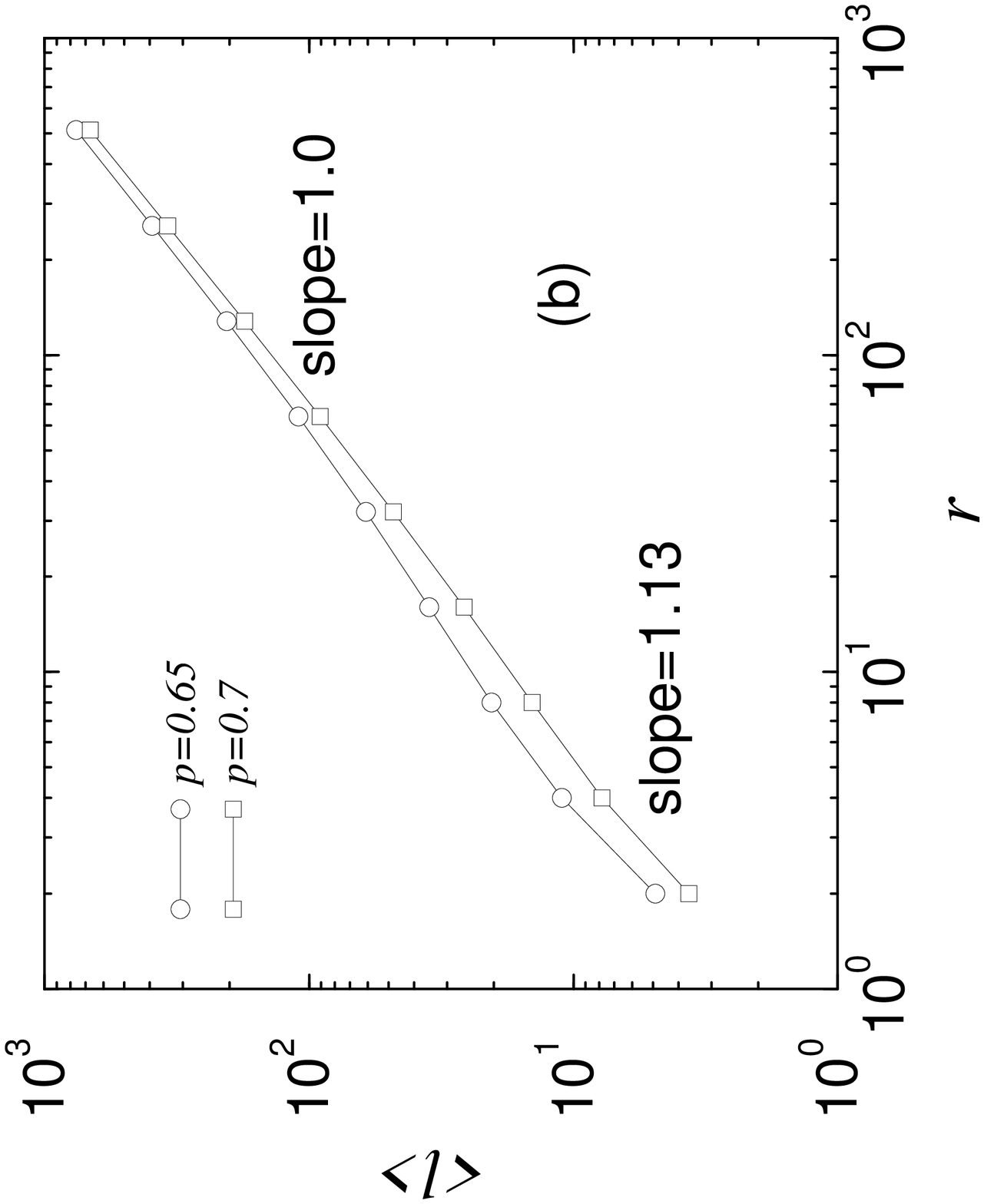}}
}
\centerline{
\epsfxsize=8.0cm
\epsfbox{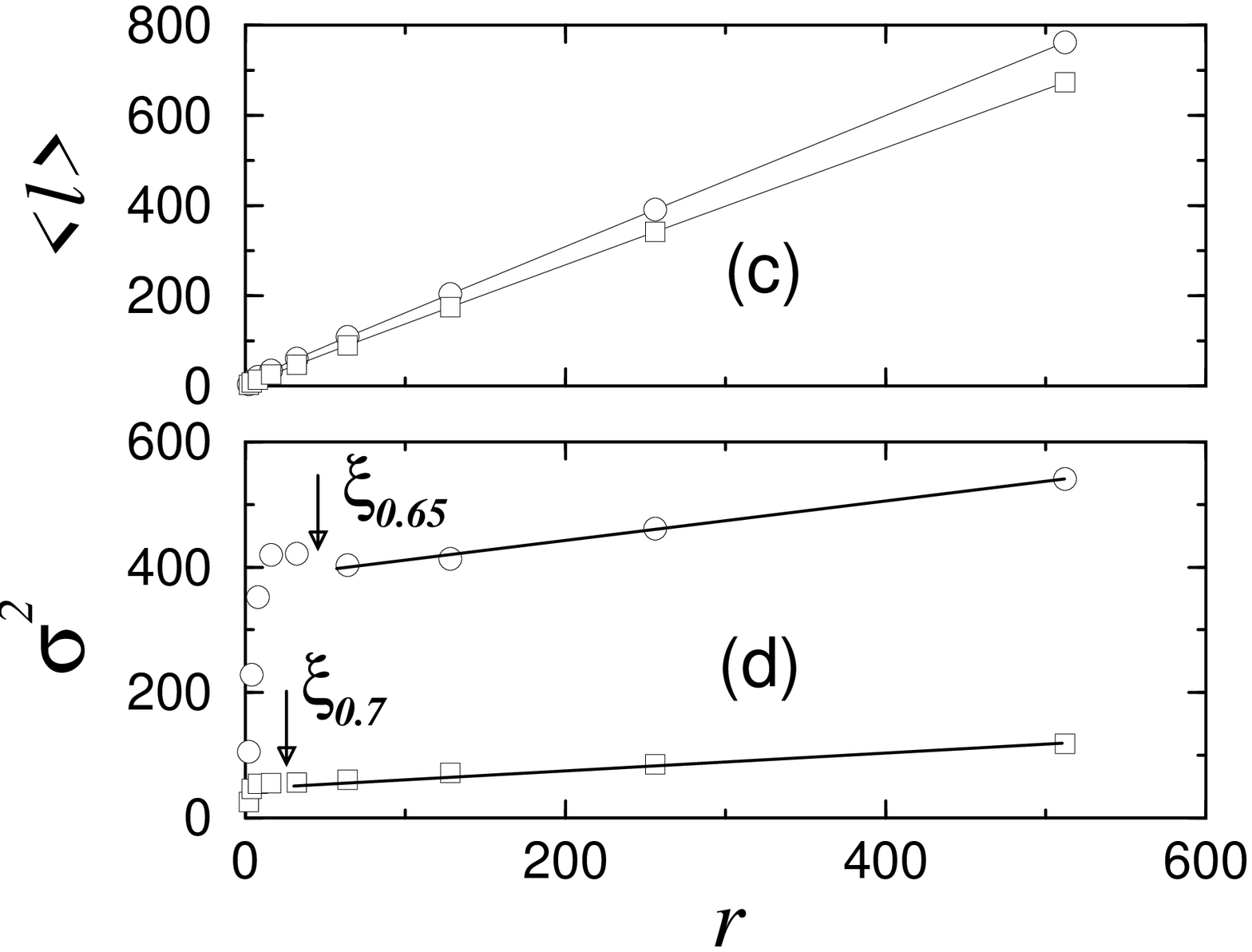}
}
\vspace{1.0cm}
\caption{
(a) Distributions of $P(\ell|r)$ for $r=4,8,16,32,64,128,256,512$ and
for $p=0.7$. To reduce the lattice effects, data is obtained for the
pairs of wells on the $x$-axis. (Note that for this case, where
$r>\xi$, the distributions $P'(\ell|r)$ and $P(\ell|r)$ are
essentially the same since all the clusters span the lattice.) The
distributions converge for large $r$ to a Gaussian with mean
$\langle\ell\rangle$ shown in parts (b,c) and variance
$\sigma^2=\langle \ell^2\rangle-\langle\ell\rangle^2$, shown in part
(d) as functions of $r$ for $p=0.65(\bigcirc)$ and $p=0.7(\Box)$. (b)
Log-log plot of $\langle \ell \rangle$ versus $r$. Note the crossover
from power law behavior with exponent
$d_{\mbox{\scriptsize{\mbox{\scriptsize min}}}}=1.13$ to linear
behavior with exponent 1.0. (c) Same as (b) on linear scale. The
slopes of linear the fits $k(p)$ are 1.45 for $p=0.65$ and 1.30 for
$p=0.7$. This yields $k(p)\sim (p-p_c)^{-0.17}$ in good agreement with
equation Eq.(\protect{\ref{e27x}}). (d) The dependence of $\sigma^2$
versus $r$. According to Eq.(\protect{\ref{e26x}}), the dependence
becomes linear only for $r>\xi\sim (p-p_c)^{-\nu}$, indicated on the
graph.  The slopes of linear fits $k(p)$ are 0.33 for $p=0.65$ and
0.12 for $p=0.7$. This gives $k(p)\sim (p-p_c)^{-1.6}$ in good
agreement with equation Eq.(\protect{\ref{e27y}}).
}
\label{fig7}
\end{figure}

\begin{figure}[htb]
\centerline{
\indent{~~~~~~~}
        \epsfxsize=6.5cm
        \rotate[r]{\epsfbox{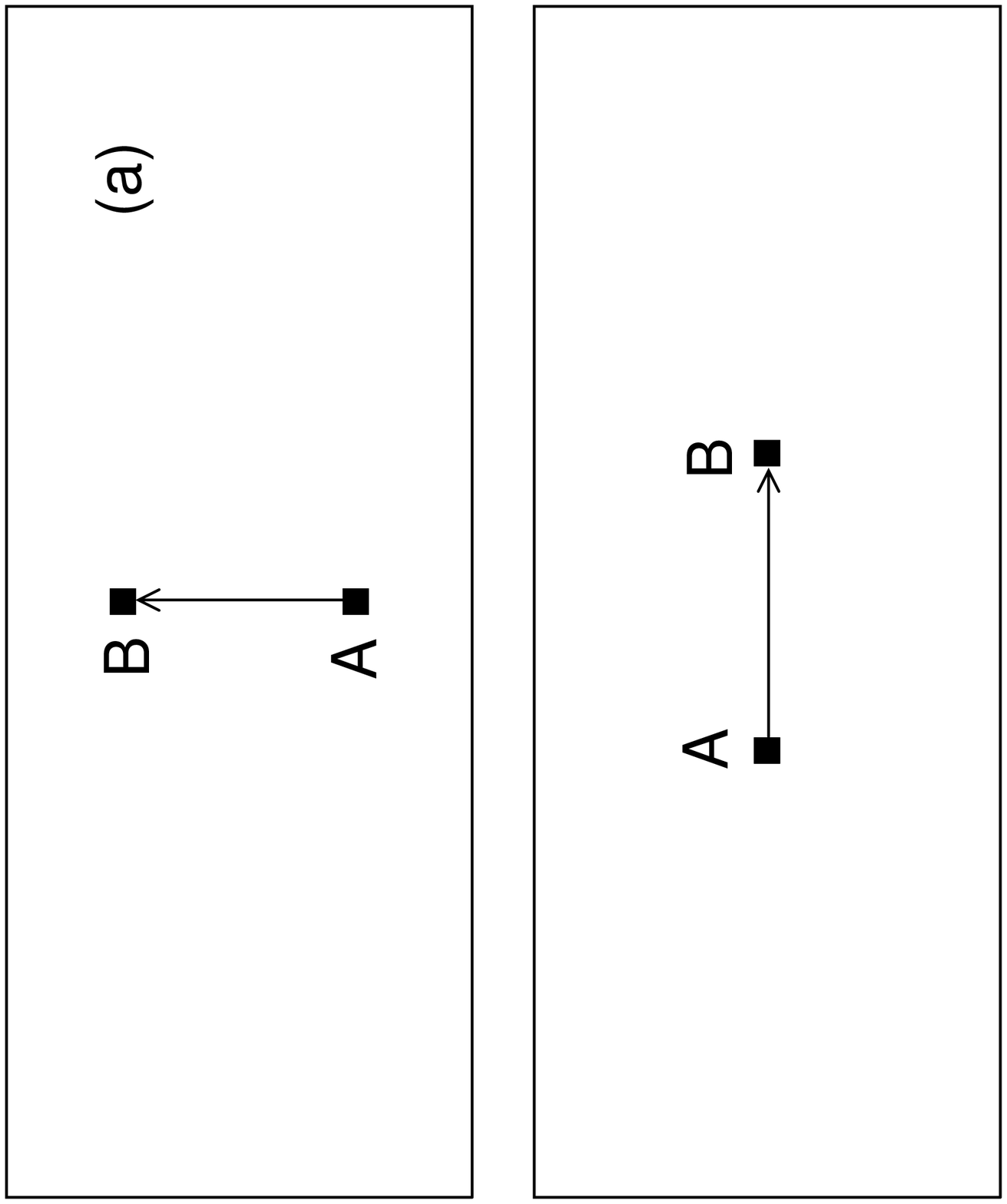}}}
\vspace*{1.0cm}
\centerline{
        \epsfxsize=8.0cm
        \rotate[r]{\epsfbox{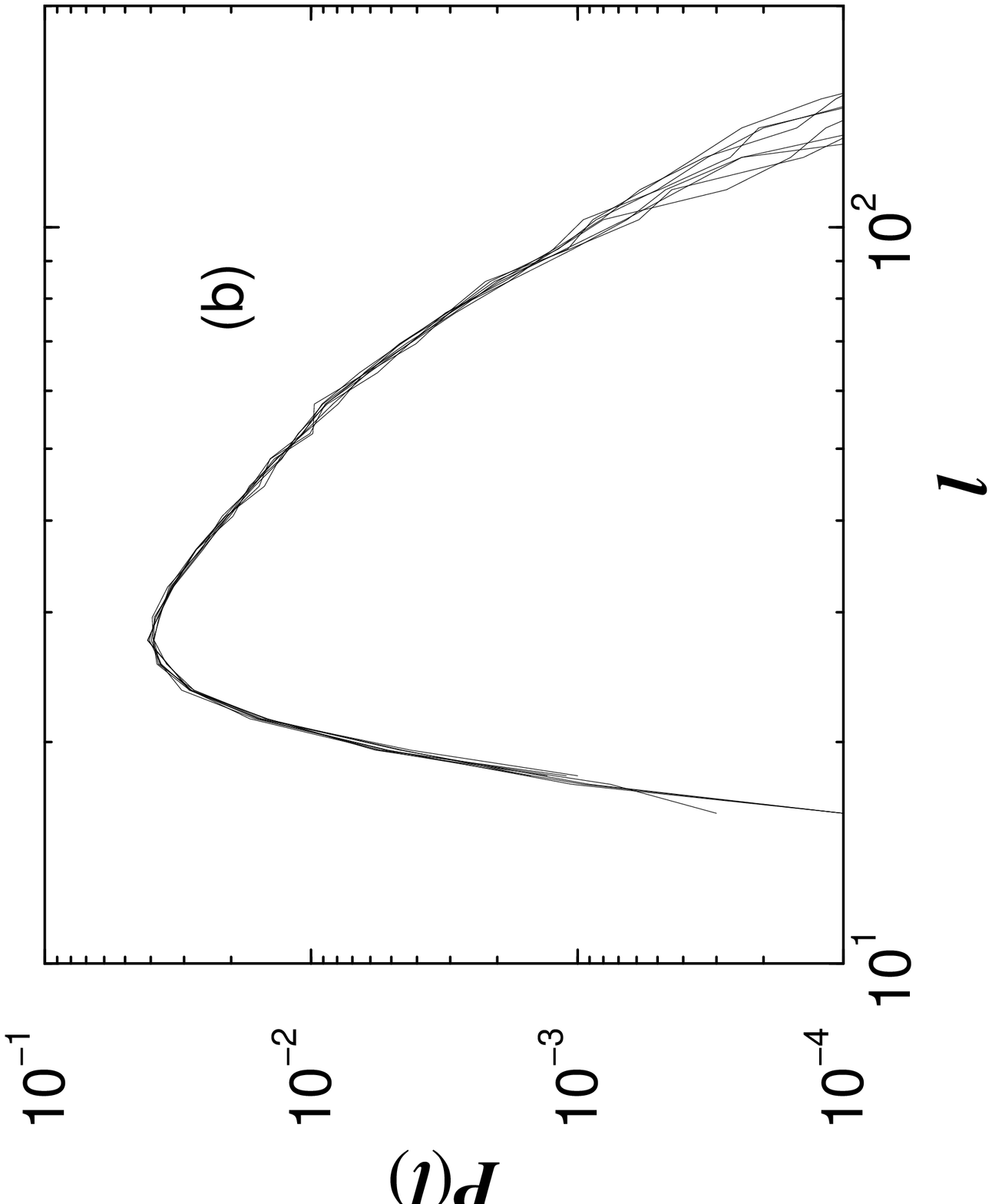}}}
\vspace*{1.0cm}
\caption{
(a) Two configurations with rectangular boundaries are studied.
(b) For $r=4$, different system sizes
$(L_x,L_y)=(32,128),(32,256),(32,512),(32,1024),(128,32),(256,32),(512,32)$
and $(1024,32)$ are studied and all the distributions of minimal path
are collapsed into each other.
}
\label{fig8}
\end{figure}

We obtain the pressure difference across each bond by solving
Kirchhoff's law
\be
\sum_j v_{ij} = 0\, ,
\label{eq:vij}
\ee
for each node $i$ in the cluster where the summation is over all bonds
connected to that node. Fig.~\ref{fig1}b shows the results of solving
these equations for the cluster discussed in Section I
(Fig.~\ref{fig1}b). Magnitudes of currents on cluster backbone are
depicted in gray scale with the lightest areas corresponding to the
smallest currents and the darkest to the largest currents.

We define the {\it traveling time}, ${\tilde t}$, of a path $\cal C$ as
the sum of the tracer's traveling times $t_{ij}$ at each bond $(ij)$
joining sites $i$ and $j$ which are on the path,
\be
\tilde t = \sum_{(ij) \in \cal C} t_{ij}.
\ee
The {\it traveling length}, ${\tilde \ell}$, in turn, is the number of
bonds present in path $\cal C$.  Among the ensemble of all paths $\{
{\cal C}\}$, we select the path $\cal C^*$ that has the {\it minimal
traveling time}, $t_{\mbox{\scriptsize min}}$,
\be
t_{\mbox{\scriptsize min}}({\cal C^*}) = \min _{\bf \{ C \} } {\tilde
t}({\cal C})\, 
\ee
and we define the {\it length\/} of the fastest path,
${\ell_{\mbox{\scriptsize min}}}$, corresponding to the minimal
traveling time, as the number of bonds present in path $\cal C^*$.  The
first quantity $t_{\mbox{\scriptsize min}}$ is the breakthrough time of
the gas/liquid that displaces the oil during recovery and has
fundamental importance to the oil industry. The quantity $\tilde{t}$
determines post-breakthrough behavior. We also define the exponents
$d_x$, where $x$ denotes ${\ell}_{\mbox{\scriptsize min}}$,
$t_{\mbox{\scriptsize min}}$, $\tilde{\ell}$ or $\tilde{t}$ by
\be
x^* \sim r^{d_x}
\en
and where $x^*$ is the characteristic (most probable) length or time of
the corresponding distribution.

Using a ``burning'' algorithm \cite{Herrmann84x}, we then find the
minimal time and the fastest path for the particle to travel between
points $A$ and $B$. Figure~\ref{fig1}c shows the propagation of the
tracer particles through the same backbone shown in Fig.~\ref{fig1}a and
\ref{fig1}b. At $t=t_{\mbox{\scriptsize min}}$, the tracer particles
spread over $t=t_{\mbox{\scriptsize min}}\cdot J$ bonds, which
constitute a subset of the backbone with fractal dimension
$d_{\mbox{\scriptsize tm}}$, which is larger than the fractal dimension
of the minimal path but smaller than the fractal dimension of the entire
backbone $d_{\mbox{\scriptsize B}}$. Hence
\begin{equation}
\label{e36xx}
d_{\mbox{\scriptsize min}}<d_{\mbox{\scriptsize
tm}}<d_{\mbox{\scriptsize B}}.
\end{equation}

\subsection{Minimal Traveling Time}
\label{ss:mtm}

\begin{figure}[htb]
\centerline{
\epsfxsize=12.0cm
\rotate[r]{\epsfbox{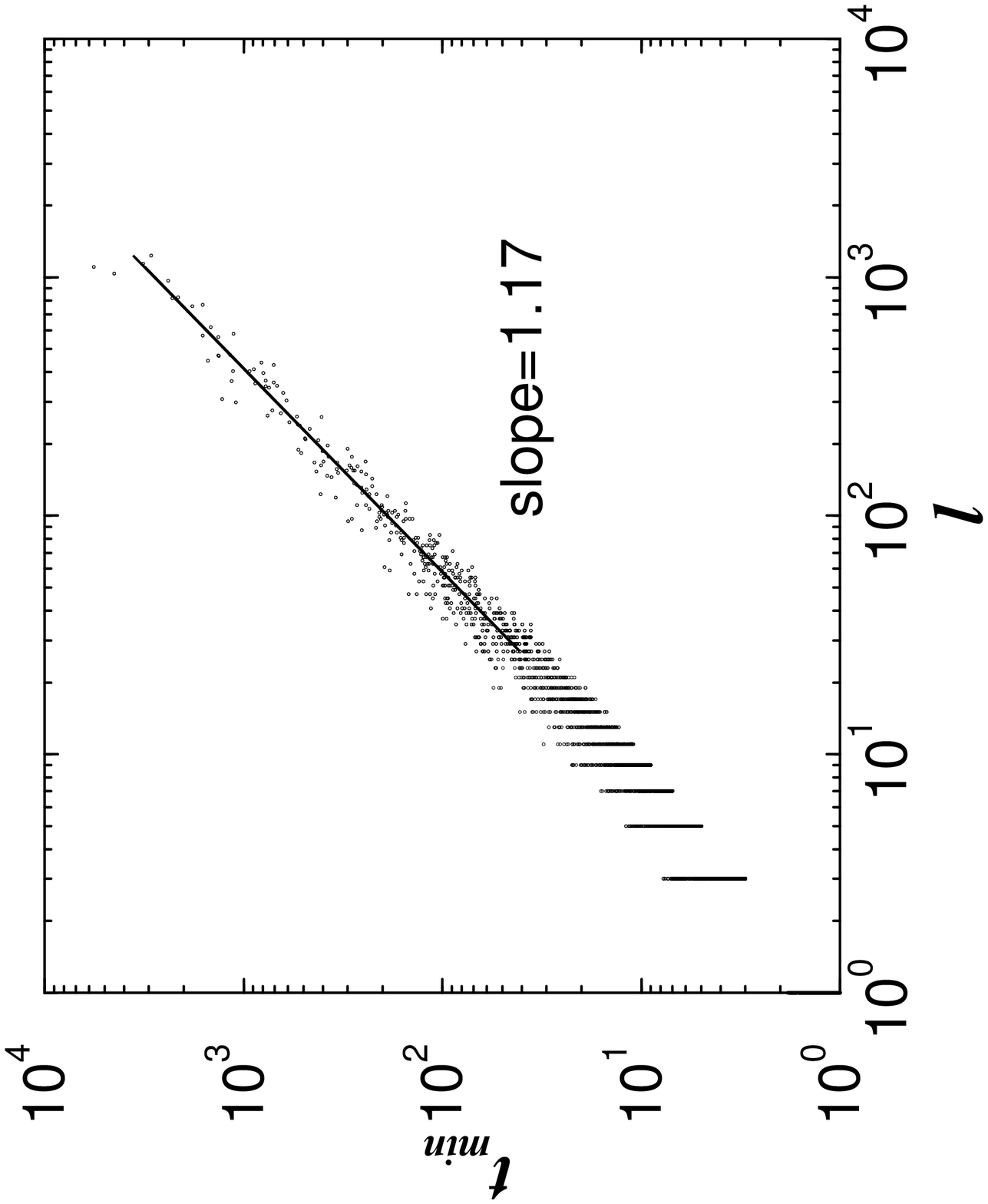}}
\vspace*{1.0cm}
}
\caption{
For $d=2$, scatter plot of the minimal traveling time
$t_{\mbox{\scriptsize min}}$ versus minimal length $\ell$ and for a
fixed well separation $r=1$. Note the strong correlation between
$t_{\mbox{\scriptsize min}}$ and $\ell$. The slope of the tail of the
scatter plot is $1.17$ yielding a value of $d_{tm}=1.17\cdot
d_{\mbox{\scriptsize min}}=1.32$, consistent with our result in Table I.
}
\label{fig10}
\end{figure}

We first study the minimal traveling time for $d=2$.  In
Fig.~\ref{fig10}, a scatter plot of the minimal traveling time versus
shortest path, we see that the minimal times are strongly correlated
with the shortest paths in the realizations simulated,
$t_{\mbox{\scriptsize min}}\sim\ell^z$, where $z\approx 1.17$. Since
$\ell$ scales as $r^{d_{\mbox{\scriptsize min}}}$ we propose that
$t_{\mbox{\scriptsize min}}$ scales as $r^{d_{\mbox{\scriptsize tm}}}$
with $d_{\mbox{\scriptsize tm}}=zd_{\mbox{\scriptsize min}}=1.33$.  This
suggests that the same scaling form which applies to the distribution of
shortest paths can also be applied to the distribution of minimal times,
but with different exponents and amplitudes.  Thus, we expect the {\it
Ansatz\/} similar to Eq.~(\ref{eq:minansatz})
\begin{eqnarray}
\nonumber P'(t_{\mbox{\scriptsize min}}|r) &\sim& {1 \over r^{d_{\mbox{\scriptsize
tm}}}}\left({t_{\mbox{\scriptsize min}} 
\over r^{d_{\mbox{\scriptsize tm}}}}\right)^{ -g_{\mbox{\scriptsize tm}}'}
f_1\left({t_{\mbox{\scriptsize min}} \over r^{d_{\mbox{\scriptsize
tm}}}}\right)\cdot\\
&~& f_2\left({t_{\mbox{\scriptsize min}} \over
L^{d_{\mbox{\scriptsize tm}}}}\right)~ f_3\left({t_{\mbox{\scriptsize
min}} \over \xi^{d_{\mbox{\scriptsize tm}}}}\right)\, ,
\label{eq:mintimeansatz}
\end{eqnarray}
where the scaling functions are $f_1(x)=\exp(-a_{\mbox{\scriptsize
tm}}x^{-\phi_{\mbox{\scriptsize tm}}})$, $f_2(x)
=\exp(-b_{\mbox{\scriptsize tm}}x^{\psi_{\mbox{\scriptsize tm}}})$ and
$f_3(x)=\exp(-c_{\mbox{\scriptsize tm}}x^{\pi_{\mbox{\scriptsize
tm}}})$.  Here $\xi$ is a characteristic length of the pair
connectedness function and has a power-law dependence on the occupancy
probability $p$ as
\be \xi \sim
|p-p_c|^{-\nu}.
\label{eq:corr} 
\ee 
The first function $f_1$ accounts for the lower cut-off due to the
constraint $\ell>r$, while $f_2$ and $f_3$ account for the upper
cut-offs due to the finite size effect and due to the finite connectedness
length, respectively.  Either $f_2$ and $f_3$ becomes irrelevant,
depending on which of the two values $L$ or $\xi$ is greater. For
$L<\xi$, $f_2$ dominates the upper cut-off, otherwise $f_3$
dominates. We have assumed independence of the finite size effect and
off-criticality effect, so that Eq.~(\ref{eq:mintimeansatz}) can be
represented as a product of the terms which are responsible for the
finite size effect ($f_2$) and the effect of the concentration ($f_3$).

We sample over $10^6$ different realizations with the two sites $A$ and
$B$ fixed. For each realization, we calculate exactly the minimal
traveling time and the path which corresponds to the minimal traveling
time to obtain $P(t_{\mbox{\scriptsize min}})$ and
$P(\ell_{\mbox{\scriptsize min}})$.

\subsubsection{Behavior at criticality}

\begin{figure}[htb]
\centerline{
\epsfxsize=7.0cm
\rotate[r]{\epsfbox{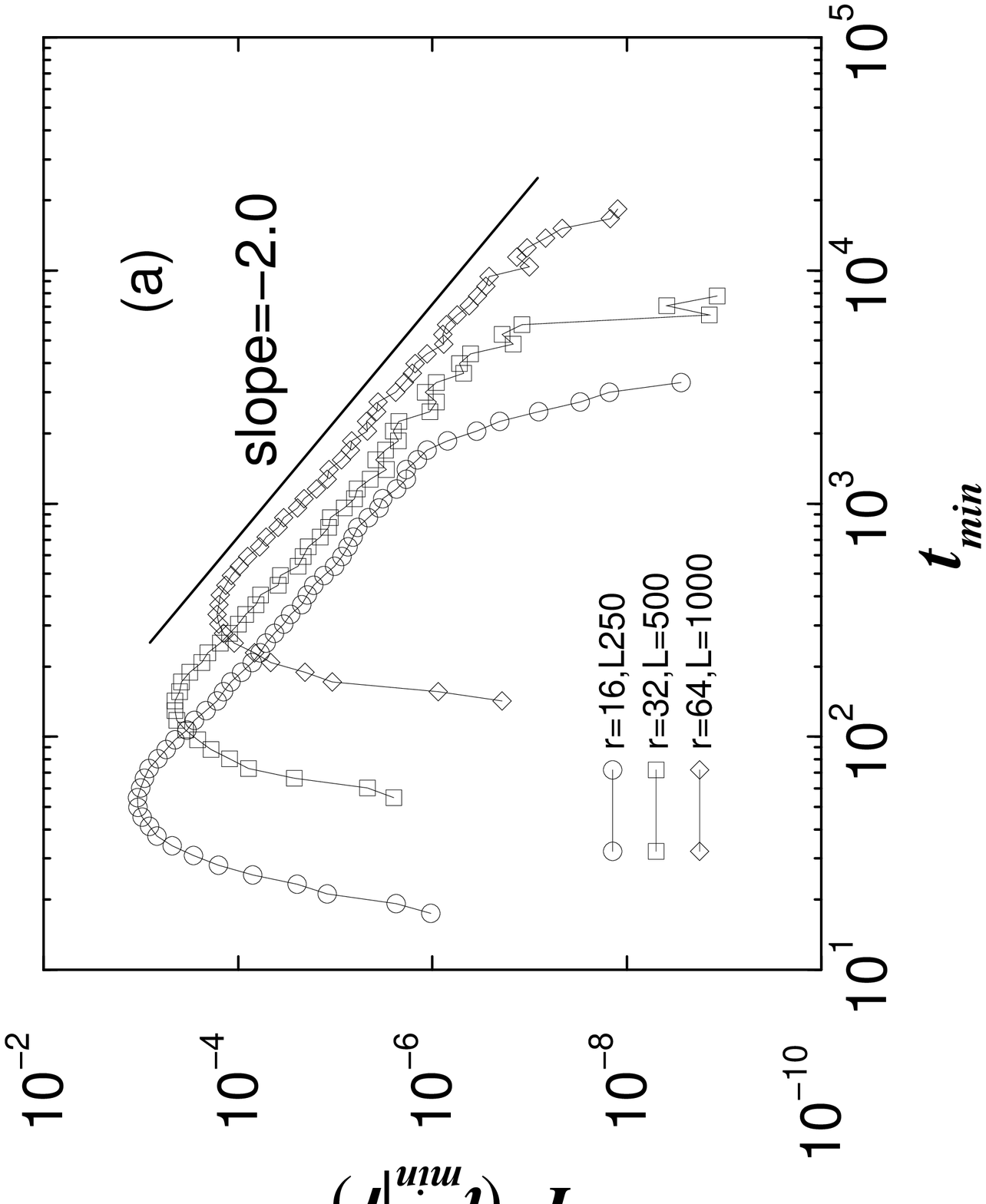}}
}
\centerline{
\epsfxsize=7.0cm
\rotate[r]{\epsfbox{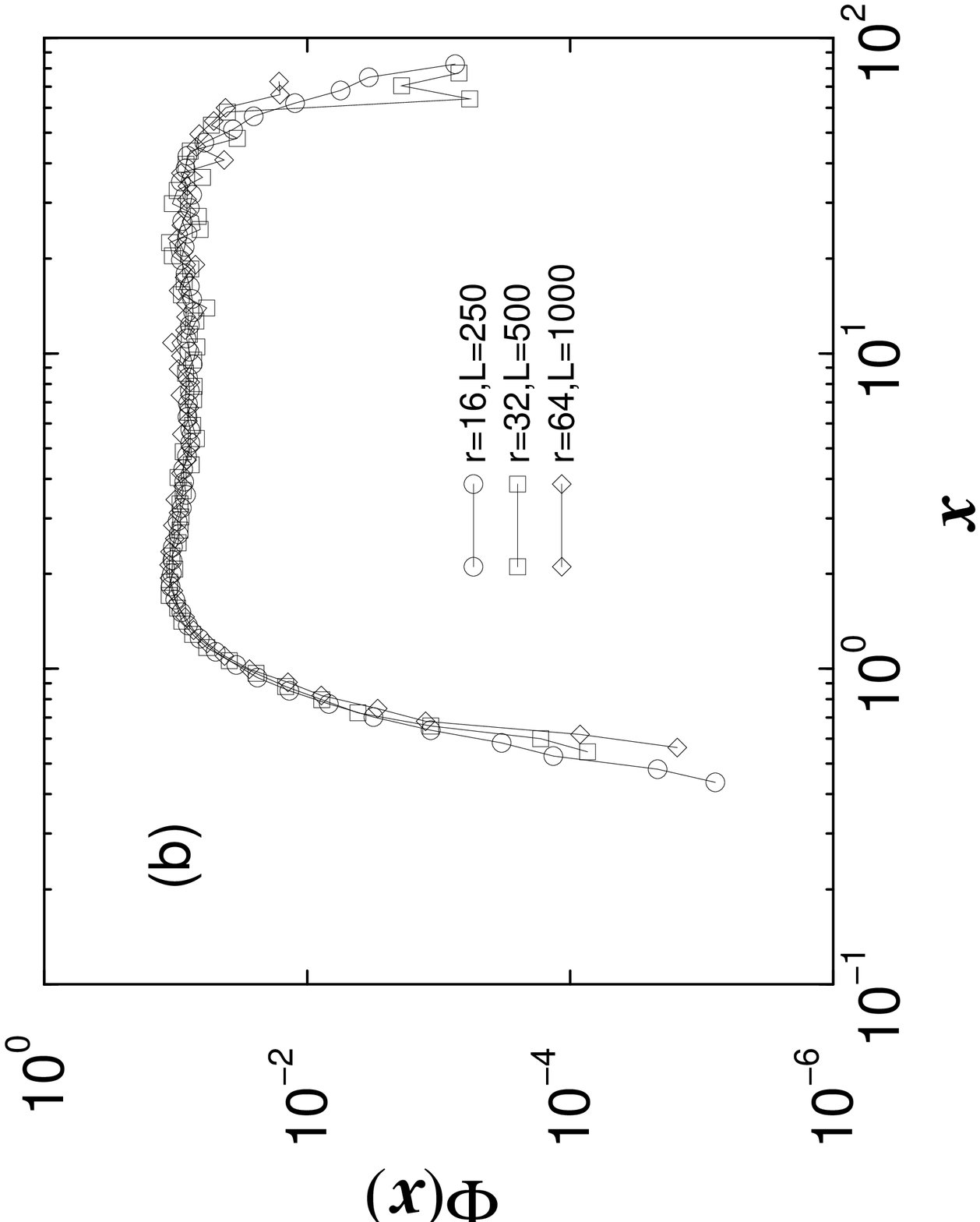}}
}
\centerline{
\epsfxsize=7.0cm
\rotate[r]{\epsfbox{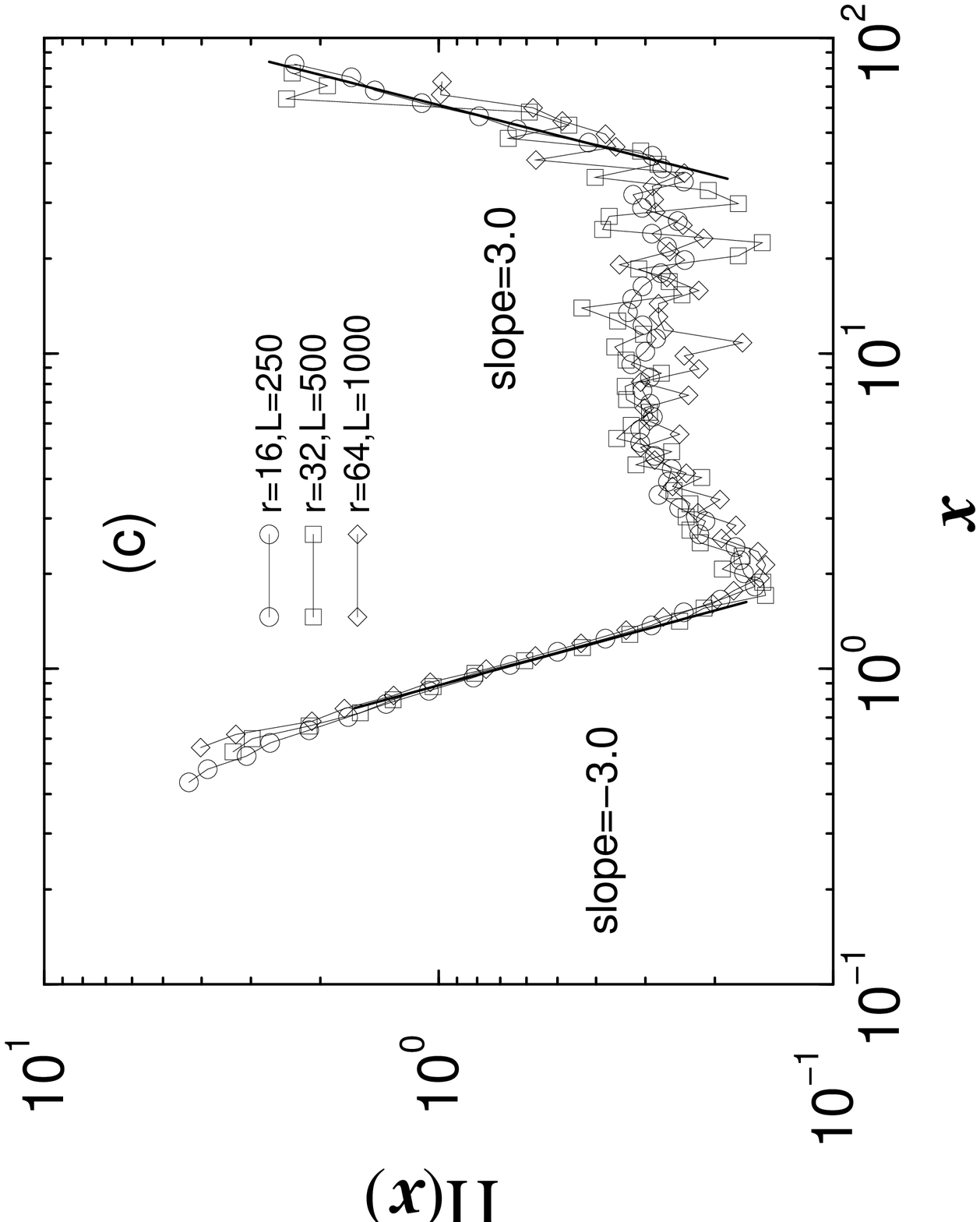}}
\vspace*{1.0cm}
}
\caption{
For $d=2$, (a) log-log plot of $P'(t|r)$ for $p=p_c=0.5$ and for
different sets of parameters, $(r,L)=(16,250),(32,500),(64,1000)$. The
straight line regime has slope $g_{t}'=2.0$. (b) Log-log plot of
rescaled probability $\Phi(x)\equiv P'(t_{\mbox{\scriptsize
min}}|r)x^{g_t'}r^{d_t}$ against rescaled length $x=t_{\mbox{\scriptsize
min}}/r^{d_t}$ using the values, $g_t'=2.0$ and $d_t=1.33$. The curves
are flat in the center because $f_2(x)$ is stretched exponential (see
Eq.~(\protect{\ref{eq:Phix}})). (c) Log-log plot of transformed
probability $\Pi(x)=\log_{10}[A/\Phi(x)]$ versus $x=t_{\mbox{\scriptsize
min}}/r^{d_t}$. The slopes of the solid lines give the power of the
stretched exponential function $f_1$ and $f_2$ in
Eq.~(\protect{\ref{eq:Phix}}). Using the parameter $A=0.14$, the slopes
give $\phi \approx 3.0$ for the lower cut-off and $\psi \approx 3.0$ for
the upper cut-off.
}
\label{fig11}
\end{figure}

We first test numerically the scaling conjecture
Eq.~(\ref{eq:mintimeansatz}) at the percolation threshold $p=p_c$. In
this case, $\xi=\infty$ and $f_3$ is a constant. Hence
Eq.~(\ref{eq:mintimeansatz}) reduces to
\begin{eqnarray}
\nonumber P'(t_{\mbox{\scriptsize min}}|r) &\sim& {1 \over r^{d_{\mbox{\scriptsize
tm}}}}\left({t_{\mbox{\scriptsize min}} \over r^{d_{\mbox{\scriptsize
tm}}}}\right)^{ -g_{\mbox{\scriptsize tm}}'}\cdot\\
&~& f_1\left({t_{\mbox{\scriptsize min}} \over r^{d_{\mbox{\scriptsize
tm}}}}\right)~ f_2\left({t_{\mbox{\scriptsize min}} \over
L^{d_{\mbox{\scriptsize tm}}}}\right) \quad (p=p_c).
\label{eq:mintimeansatz1}
\end{eqnarray}
Figure~\ref{fig11}a shows that $P'(t_{\mbox{\scriptsize min}}|r)$ has a power-law regime
with slope 
\begin{equation}
\label{e28a}
g_{\mbox{\scriptsize tm}}'=2.0 \pm 0.1. 
\end{equation}

To determine the functions $f_1$ and $f_2$, we compute the rescaled
probability distribution
\be 
\Phi\left({t_{\mbox{\scriptsize min}}\over
r^{d_{\mbox{\scriptsize tm}}}}\right) \equiv P'(t_{\mbox{\scriptsize
min}}|r)(t_{\mbox{\scriptsize min}})^{g_{\mbox{\scriptsize
tm}}'}r^{-d_{\mbox{\scriptsize
tm}}(g_{\mbox{\scriptsize
tm}}-1)},
\label{eq:trescaled} 
\en
and plot it against scaling variable $x=t_{\mbox{\scriptsize
min}}/r^{d_{\mbox{\scriptsize tm}}}$ (see Fig.~\ref{fig11}b). According
to Eq.~(\ref{eq:mintimeansatz1})
\be
\Phi(x)=Af_1(x)f_2\left[ x\left({r\over L}\right)^{d_{\mbox{\scriptsize
tm}}}\right]. \label{eq:tPhix}
\en
Therefore, $\Phi(x)$ should depend only on $x$ and the ratio
$r/L$. Unlike the fractal dimension of the shortest path,
$d_{\mbox{\scriptsize min}}$, there have been no calculations of the
fractal dimension of the minimal traveling time, $d_{\mbox{\scriptsize
tm}}$. We estimate $d_{\mbox{\scriptsize tm}}$ by finding the value
which yields the best data collapse for Eq.~(\ref{eq:tPhix}). For
$d_{\mbox{\scriptsize tm}}=1.33$, Fig.~\ref{fig11}b shows excellent data
collapse with sharp cutoffs governed for small $x<1$ by $f_1(x)$ and for
large $x>(L/r)^{d_{\mbox{\scriptsize tm}}}$ by
$f_2[x(r/L)^{d_{\mbox{\scriptsize tm}}}]$.

In order to test the assumption that the functions $f_1$ and $f_2$ are
stretched exponentials with exponents $\phi_{\mbox{\scriptsize tm}}$ and
$\psi_{\mbox{\scriptsize tm}}$, we make a log-log plot of
$\Pi(x) \equiv \log_{10}[A/\Phi(x)]$ versus $x$ for various values of the
normalization constant $A$ (See Fig.~\ref{fig11}c). If the stretched
exponential conjecture is correct, $\Pi(x)$ should have two straight
line asymptotes for $\log x \rightarrow +\infty$ with the slope
$\psi_{\mbox{\scriptsize tm}}$ and for $\log x \rightarrow -\infty$ with
the slope $-\phi_{\mbox{\scriptsize tm}}$.  The slopes
$\phi_{\mbox{\scriptsize tm}}$ and $\psi_{\mbox{\scriptsize tm}}$ of the
straight line fits depend weakly on the value of $A$.  Using $A=0.14$,
we obtain the longest regimes of straight line behavior. For this $A$ we
obtained $\phi_{\mbox{\scriptsize tm}} \approx 3.0$ and
$\psi_{\mbox{\scriptsize tm}}\approx 3.0$.  With the same assumptions
used to derive Eq.~(\ref{eq:phi_l}), we can derive a similar expression
for $\phi_{\mbox{\scriptsize tm}}$
\be
\phi_{\mbox{\scriptsize tm}}={1 \over d_{\mbox{\scriptsize tm}}-1 },
\label{eq:phi_t}
\ee
which yields a predicted value of $\phi_{\mbox{\scriptsize tm}}$ of
$3.0$ in good agreement with our simulation result.

\subsubsection{Off-Critical Behavior}

Finally, in order to test the dependence of $P'(t_{\mbox{\scriptsize
min}}|r)$ on $p$ we obtain data for large system size $L$ ($L=1000$) and
for several values of $p\neq p_c$. As we do for the shortest length, we
analyze the behavior of $t_{\mbox{\scriptsize min}}$ in three regimes
determined by the relation of the value of the connectedness length,
$\xi$, to the values of $r$ and $L$.

\begin{figure}[htb]
\centerline{
\epsfxsize=7.0cm
\rotate[r]{\epsfbox{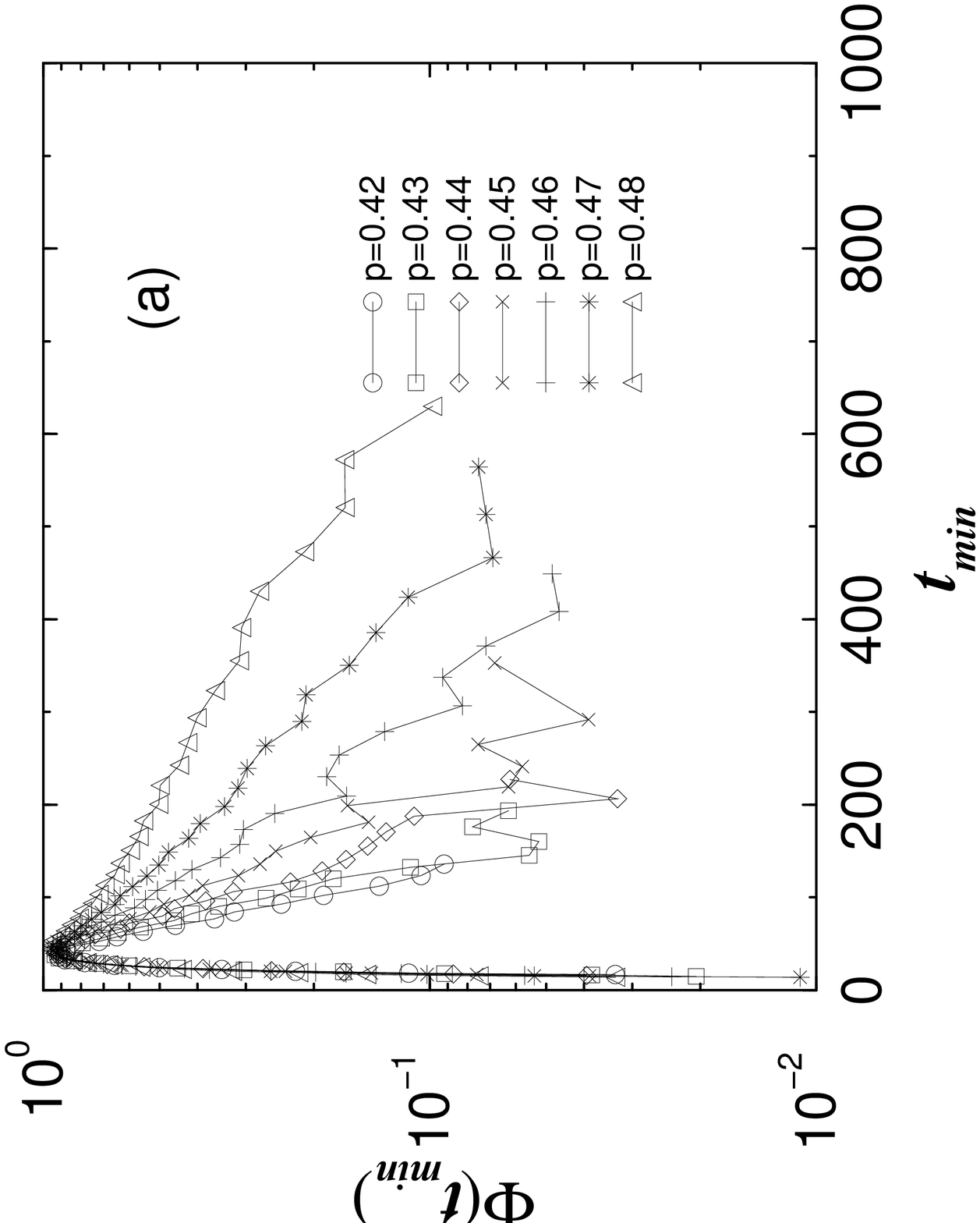}}
}
\centerline{
\epsfxsize=7.0cm
\rotate[r]{\epsfbox{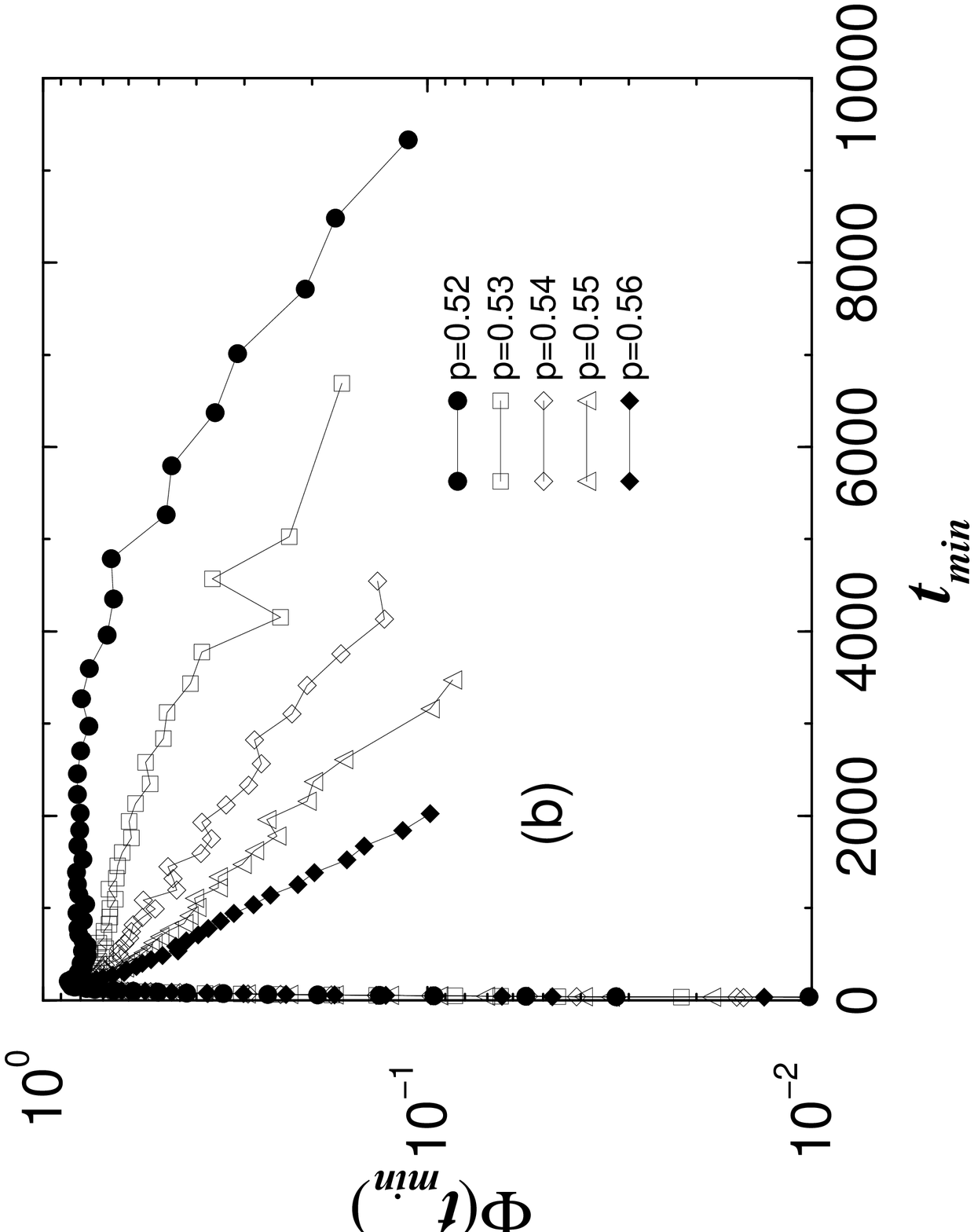}}
}
\centerline{
\epsfxsize=7.0cm
\rotate[r]{\epsfbox{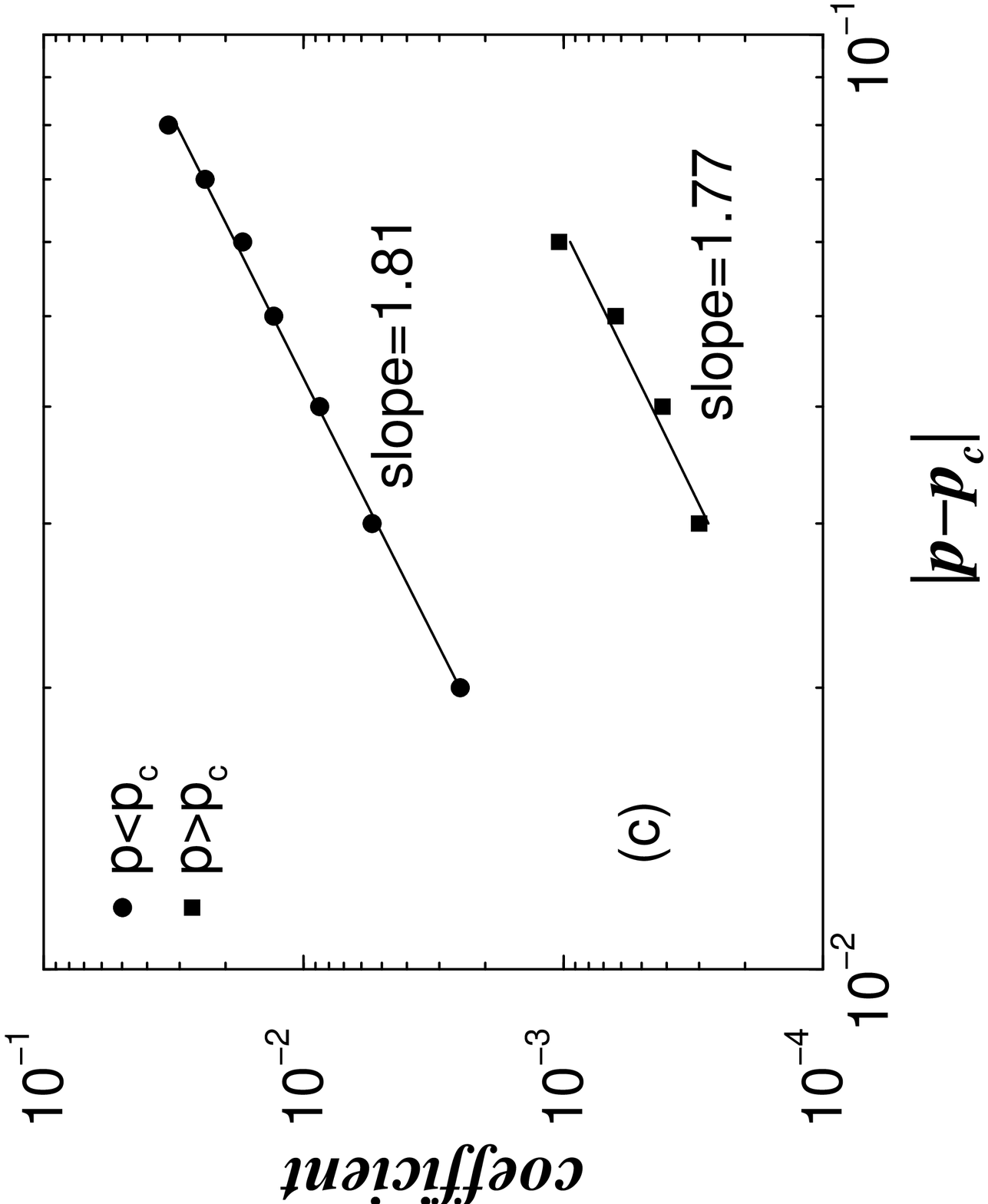}}
}
\vspace{1.0cm}
\caption{
For $d=2$, (a) semi-logarithmic plot of transformed probability
$\Phi(t_{\mbox{\scriptsize min}}/r^{d_t})$ versus $t_{\mbox{\scriptsize
min}}$ for $f_3$ for $p=0.42,0.43,0.44,0.45,0.46,0.47,0.48$ below
criticality. (b) Same for $p=0.52,0.53,0.54,0.55,0.56$ above
criticality. (c) The slope of the log-log plot of the coefficient in
exponential function $f_3$ as a function of $|p-p_c|$ gives the value
$\nu d_t \approx 1.77$ for $p>p_c$ and $1.81$ for $p<p_c$.
}
\label{fig12}
\end{figure}

\begin{figure}[htb]
\centerline{
\epsfxsize=7.0cm
\rotate[r]{\epsfbox{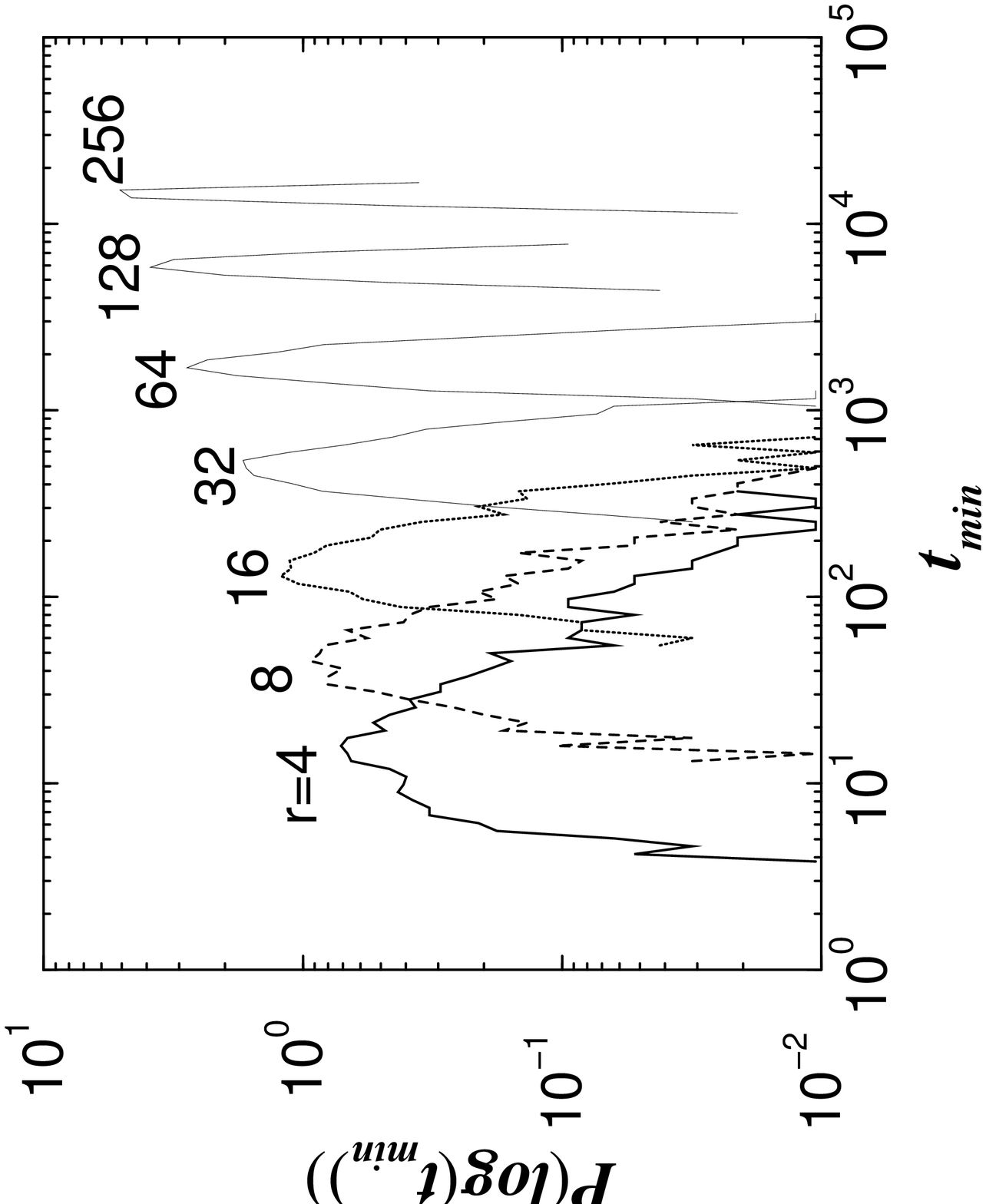}}
}
\centerline{
\epsfxsize=7.0cm
\rotate[r]{\epsfbox{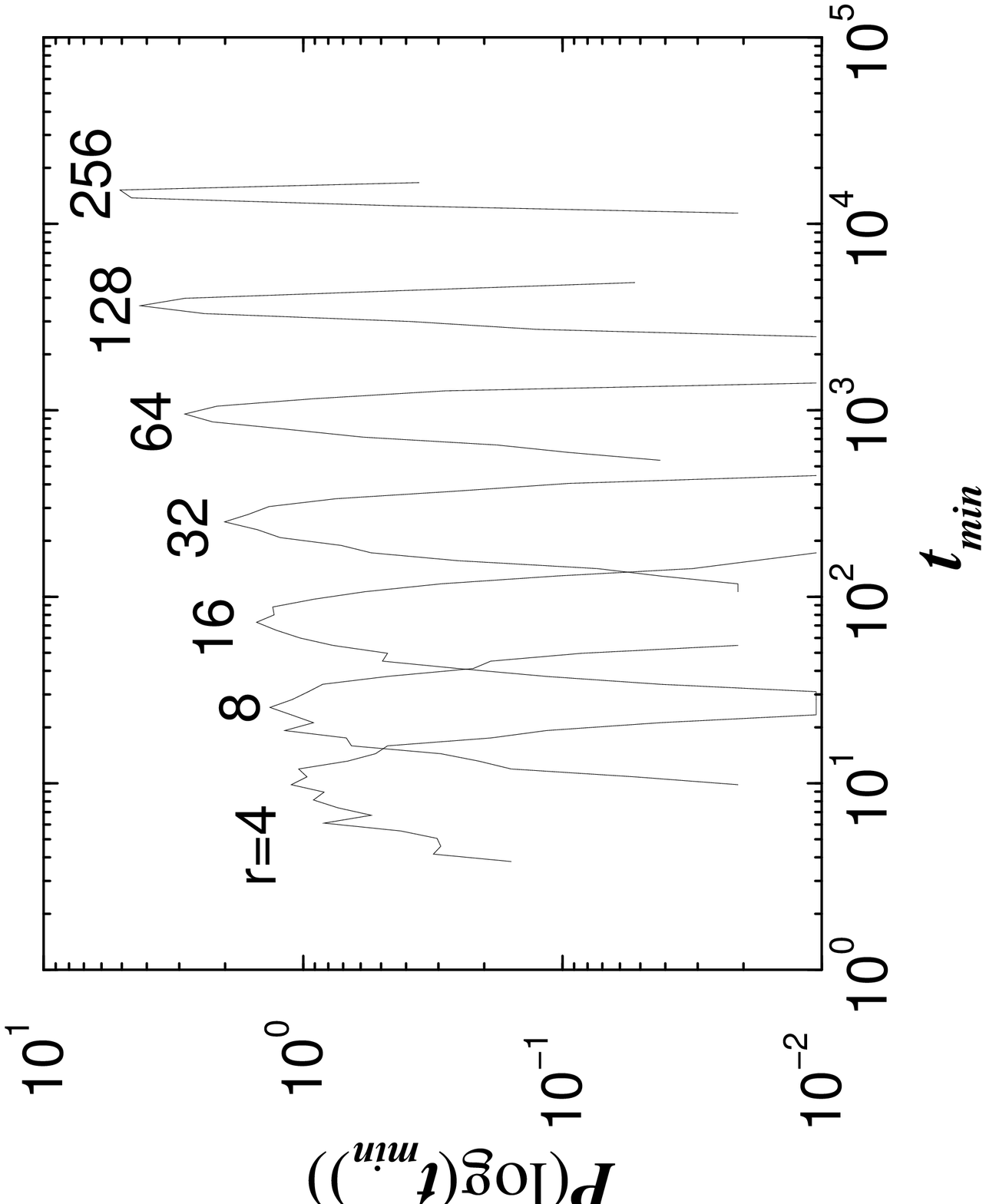}}
}
\centerline{
\epsfxsize=7.0cm
\rotate[r]{\epsfbox{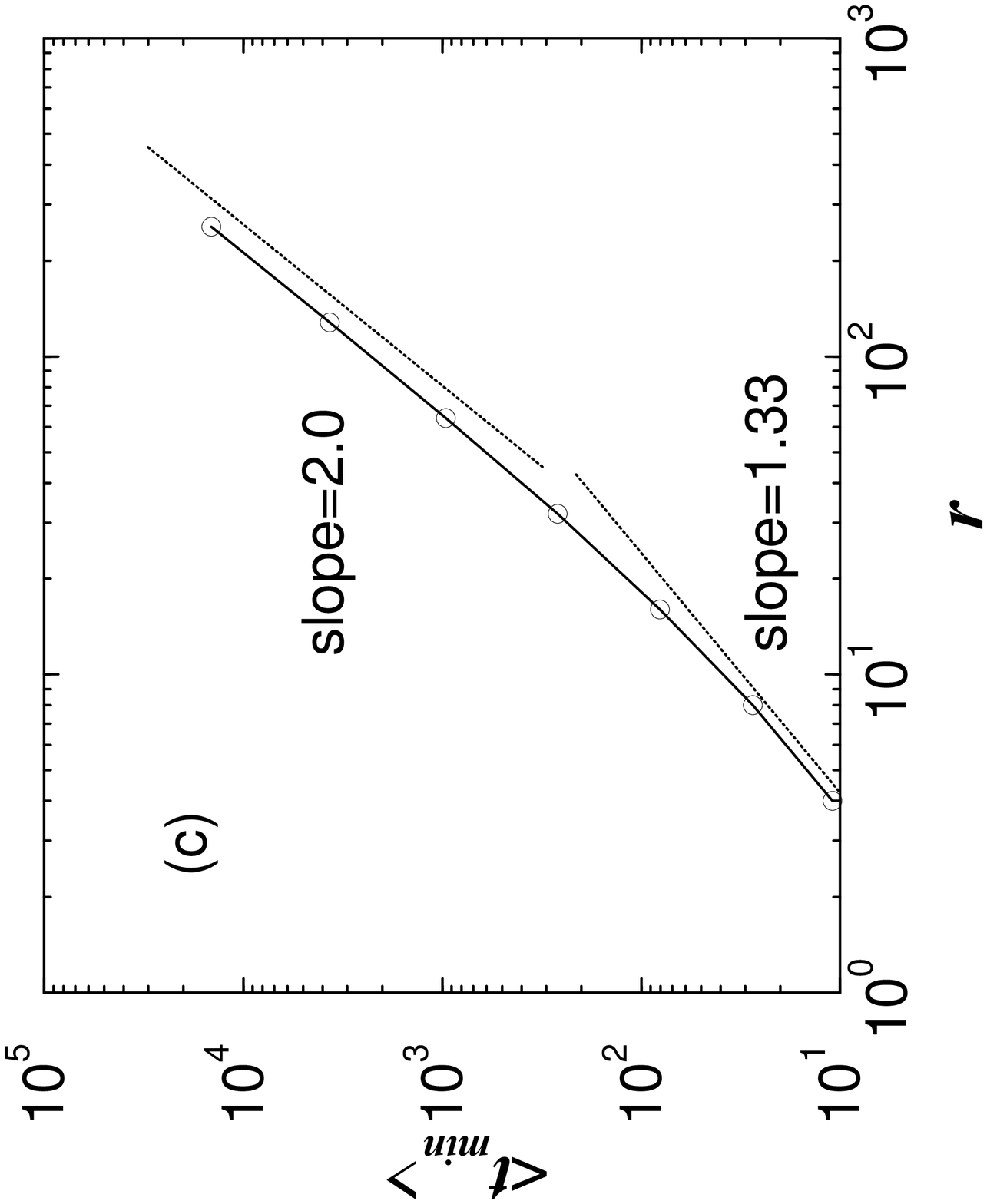}}
}
\centerline{
\epsfxsize=7.0cm
\rotate[r]{\epsfbox{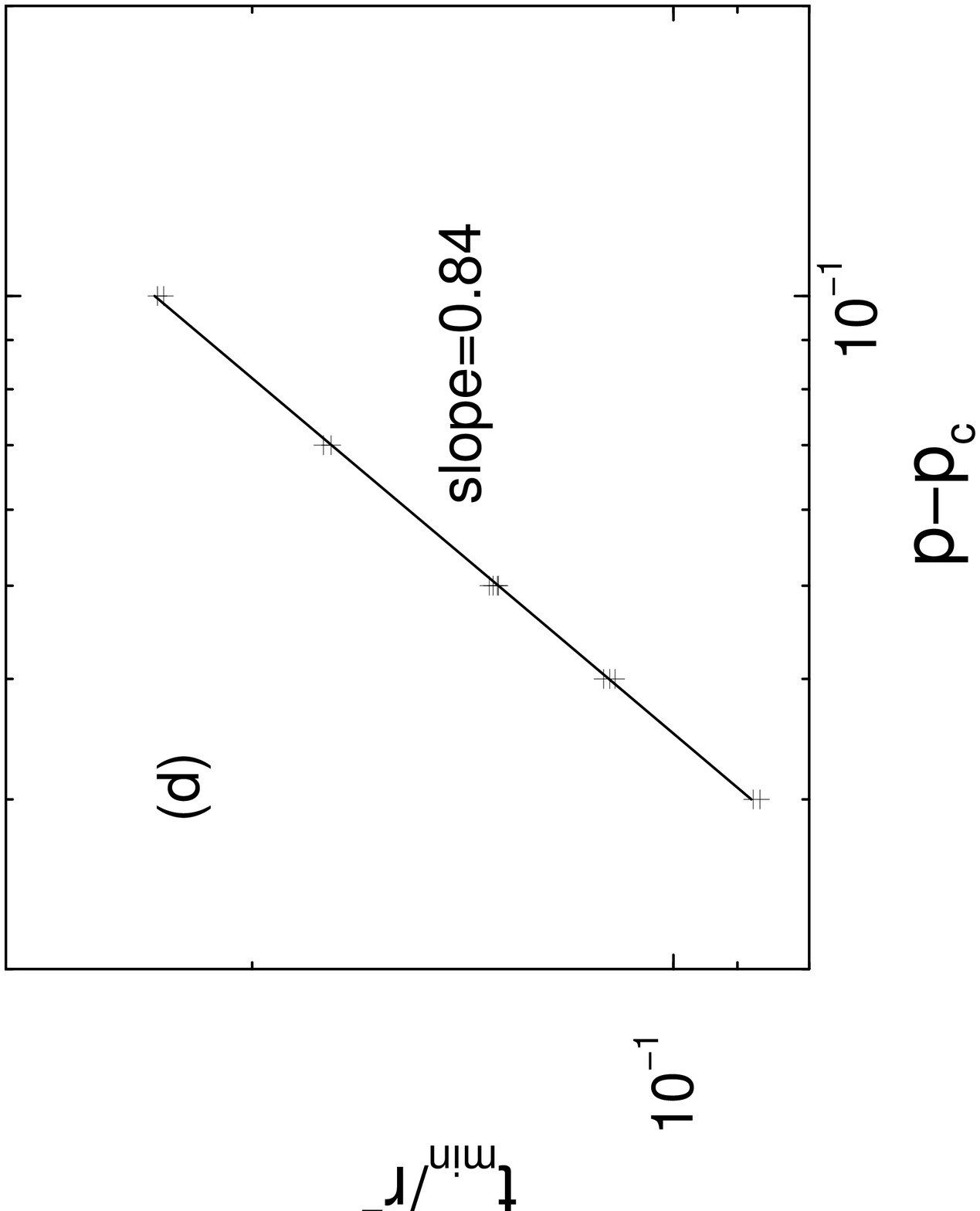}}
}
\centerline{
\epsfxsize=7.0cm
\rotate[r]{\epsfbox{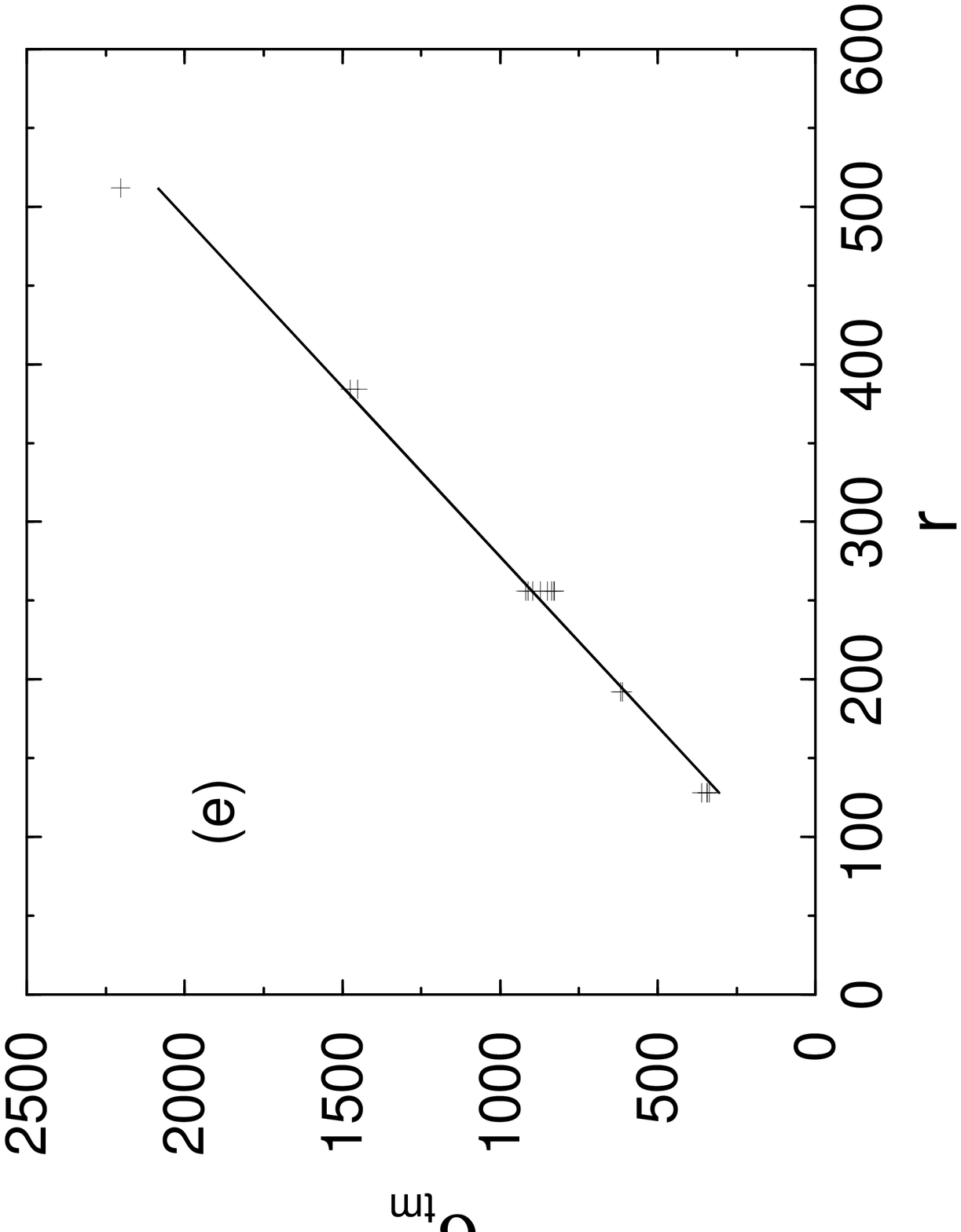}}
}
\vspace{1.0cm}
\caption{
(a) Log-log plot of $P(t_{\mbox{\scriptsize min}}|r)$ for $p=0.6$ and
for $r=4,8,16,32,64,128,256$ and $L=258$. The distributions for large
$r$ converge to Gaussians with mean $\langle t_{\mbox{\scriptsize
min}}\rangle$ and variance $\sigma^2$. (b) Log-log plot of
$P(t_{\mbox{\scriptsize min}}|r)$ for $p=0.6$, $r=4,8,16,32,64,128,256$
and $L=r+2$. (Note that for this case, where $r>\xi$, the distributions
$P'(t_{\mbox{\scriptsize min}}|r)$ and $P(t_{\mbox{\scriptsize min}}|r)$
are essentially the same since all the clusters span the lattice.) (c)
Log-log plots of $\langle t_{\mbox{\scriptsize min}}\rangle$ versus $r$
for $p=0.6$ and $L=r+2$.  (d) Log-log plot of the scaled average minimal
traveling time, $\langle t_{\mbox{\scriptsize min}}\rangle/r^2$, versus
$p-p_c$ for $r=128,192,256,384,512$ and $L=r+2$. Note that in all cases
$r \gg \xi$. The slope of the line, $0.84$, is in good agreement with
the theoretical prediction, $0.89$. (e) The behavior of the width,
$\sigma$, of the distributions of the traveling time versus $r$ for
$p=0.53,0.54,0.55,0.57$ and $0.6$. The graph shows approximately linear
dependence of $\sigma$ on $r$. The variation of the slope with $p-p_c$
is within the error bars of the data.
}
\label{fig14}
\end{figure}

\begin{itemize}

\item[{(i)}] $\xi>L>r$. In this regime, the fact that $p\neq p_c$ cannot
be detected because the connectedness length is larger than the other
relevant variables.

\item[{(ii)}] $L>\xi>r$. In this case, the upper cutoff of the distribution
Eq.~(\ref{eq:mintimeansatz}) is governed by $f_3$ and the functional
form of the rescaled probability $\Phi$ is given by
\be
\Phi(t_{\mbox{\scriptsize min}}/r^{d_{\mbox{\scriptsize tm}}})
\sim f_1\left({t_{\mbox{\scriptsize min}}\over r^{d_{\mbox{\scriptsize
tm}}}}\right) 
f_3\left({t_{\mbox{\scriptsize min}}\over\xi^{d_{\mbox{\scriptsize
tm}}}}\right).
\label{eq:tphix}
\en    
For large $t_{\mbox{\scriptsize min}}$, we suggest an exponential decay
of $\Phi$
\be
\Phi(t_{\mbox{\scriptsize min}}/r^{d_{\mbox{\scriptsize tm}}})
\sim \exp\left(-c{t_{\mbox{\scriptsize
min}}\over\xi^{d_{\mbox{\scriptsize 
tm}}}}\right). 
\label{eq:tplrp}
\en
Semi-logarithmic plots of $\log \Phi(t_{\mbox{\scriptsize
min}}/r^{d_{\mbox{\scriptsize tm}}})$ versus $t_{\mbox{\scriptsize
min}}$ for $p>p_c$ and $p<p_c$ shown in Fig.~\ref{fig12}a and
\ref{fig12}b, respectively, can be approximated by straight lines with
slopes which approach zero as $p \rightarrow p_c$.  According to
Eq.~(\ref{eq:tplrp}), this slope $k(p)$ should be proportional to
\begin{equation}
\label{e34a}
\xi^{-d_{\mbox{\scriptsize tm}}}=|p-p_c|^{d_{\mbox{\scriptsize
tm}}\nu}\approx |p-p_c|^{1.77}.  
\end{equation}
Figure~\ref{fig12}c shows double logarithmic plots of $|k(p)|$ versus
$|p-p_c|$ for $p<p_c$ and $p>p_c$, which can be well approximated by
straight lines with slopes 1.81 and 1.77, respectively, in good
agreement with the scaling conjecture, Eq.~(\ref{e34a}).  As was the
case with the analysis of $P'(\ell|r)>p_c$ for $d=3$ (see
Sec.~II.C.2.ii), we cannot determine the parameters that govern the
large $t_{\mbox{\scriptsize min}}$ behavior of $P'(t_{\mbox{\scriptsize
min}})$ because of limitations on the size of the system we can
simulate.

\item[{(iii)}] $L>r>\xi$. When the connectedness length is smaller
than the distance between the wells, the behavior of the system is the
same as a homogeneous system \cite{Koplik94,King99}. This can be seen
in Fig.~\ref{fig14}a in which we plot $P(t_{\mbox{\scriptsize
min}}|r)$ for various values of $r$ at $p=0.6$. As $r$ increases from
below the connectedness length to above the connectedness length, the
form of the distribution changes from the power law distribution of
Eq.~(\ref{eq:mintimeansatz1}) to a distribution with a pronounced
peak, a characteristic of homogeneous systems. In Fig.~\ref{fig14}b,
in order to eliminate the finite-size effect, we select $L=r+2$ so
that the distribution $P(t|r)$ does not have a power-law regime, even
for small $r$.  In this case, as shown in Fig.~\ref{fig14}c, the
fractal dimension of the minimal traveling time crosses over from
$d_{\mbox{\scriptsize tm}}=1.33$ to $d_{\mbox{\scriptsize tm}}=2.0$,
characteristic of a homogeneous system \cite{Koplik94,King99}.  The
same considerations that we use to derive the behavior of the mean and
variance of the shortest path can be applied to the mean and variance
of the minimal time. At the moment of breakthrough, i.e., when the
first tracer particle reaches the second well, the part of the system
filled with tracer particles consists of $n_b=(r/\xi)^d$ independent
blobs, each of which having a certain number of bonds
$(t_{\mbox{\scriptsize min}})_b$ with an average $\langle
(t_{\mbox{\scriptsize min}})_b\rangle=\xi^{d_{\mbox{\scriptsize tm}}}$
and a variance $\sigma_b^2=\xi^{2d_{\mbox{\scriptsize tm}}}$. Thus the
average minimal time for the entire system scales as
\begin{equation}
\label{e35x}
\langle t_{\mbox{\scriptsize min}}\rangle=n_b\cdot\xi^{d_{\mbox{\scriptsize
tm}}}=r^d\xi^{d_{\mbox{\scriptsize tm}}-d},
\end{equation}
with a variance
\begin{equation}
\label{e36x}
\sigma^2=n_b\xi^{2d_{\mbox{\scriptsize tm}}}=r^d\xi^{2d_{\mbox{\scriptsize
tm}}-d}.
\end{equation}
The scaling plot (Fig.~\ref{fig14}d) of $\langle t_{\mbox{\scriptsize
min}}\rangle$ versus $|p-p_c|$ shows good agreement with the theoretical
prediction of Eq.~(\ref{e35x})
\begin{eqnarray}
\label{e36xxx}
\nonumber{\langle t_{\mbox{\scriptsize min}}\rangle\over
r^d} &=& (p-p_c)^{(d-d_{\mbox{\scriptsize tm}})\nu}\\
&=& (p-p_c)^{0.89}\qquad [d=2].
\end{eqnarray}
The graph of $\sigma$ versus $r$ (see Fig.~\ref{fig14}e) shows linear
behavior of $\sigma$ versus $r$ in agreement with
Eq.~(\ref{e36x}). Equation (\ref{e36x}) also predicts that the slope of
this linear dependence decays as
\begin{equation}
\label{e36yy}
|p-p_c|^{-[d_{\mbox{\scriptsize
tm}}-(d/2)]\nu}=|p-p_c|^{-0.42}\qquad\qquad [d=2]. 
\end{equation}
However the measured slope has very small variation with $|p-p_c|$ which
is beyond the accuracy of our data points.

\end{itemize}

As mentioned above, the minimal traveling time is the sum of the inverse
local velocities over the fastest path where the fastest path is
statistically identical to the shortest path. While the velocity
distribution has been studied extensively
\cite{Arcangelis85,Barthelemy99}, because the velocities along the path
are correlated, the relation between the minimum traveling time
distribution and the local velocity distribution is an open challenge
for further research.

\begin{figure}[htb]
\centerline{
\epsfxsize=7.0cm
\rotate[r]{\epsfbox{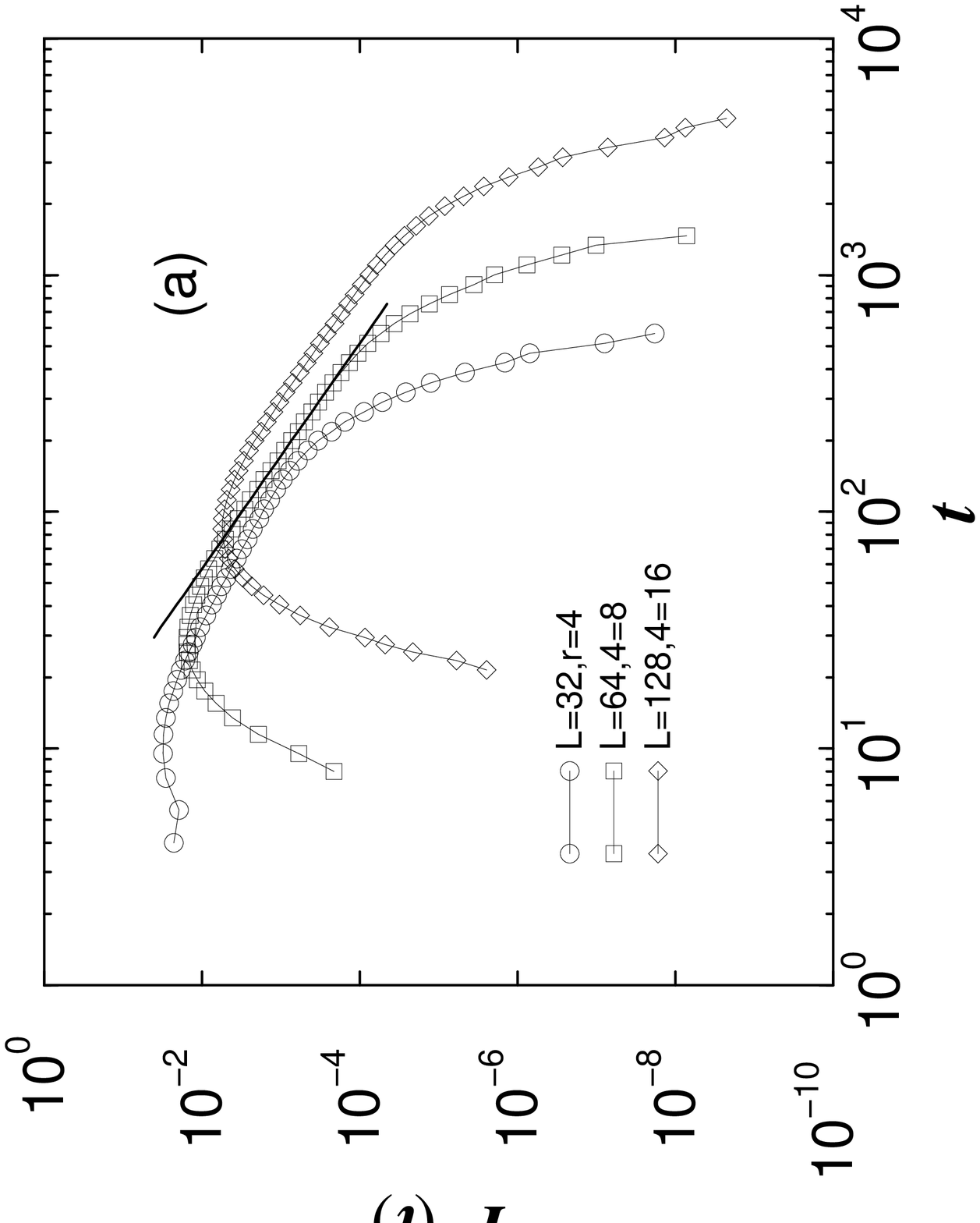}}
}
\centerline{
\epsfxsize=7.0cm
\rotate[r]{\epsfbox{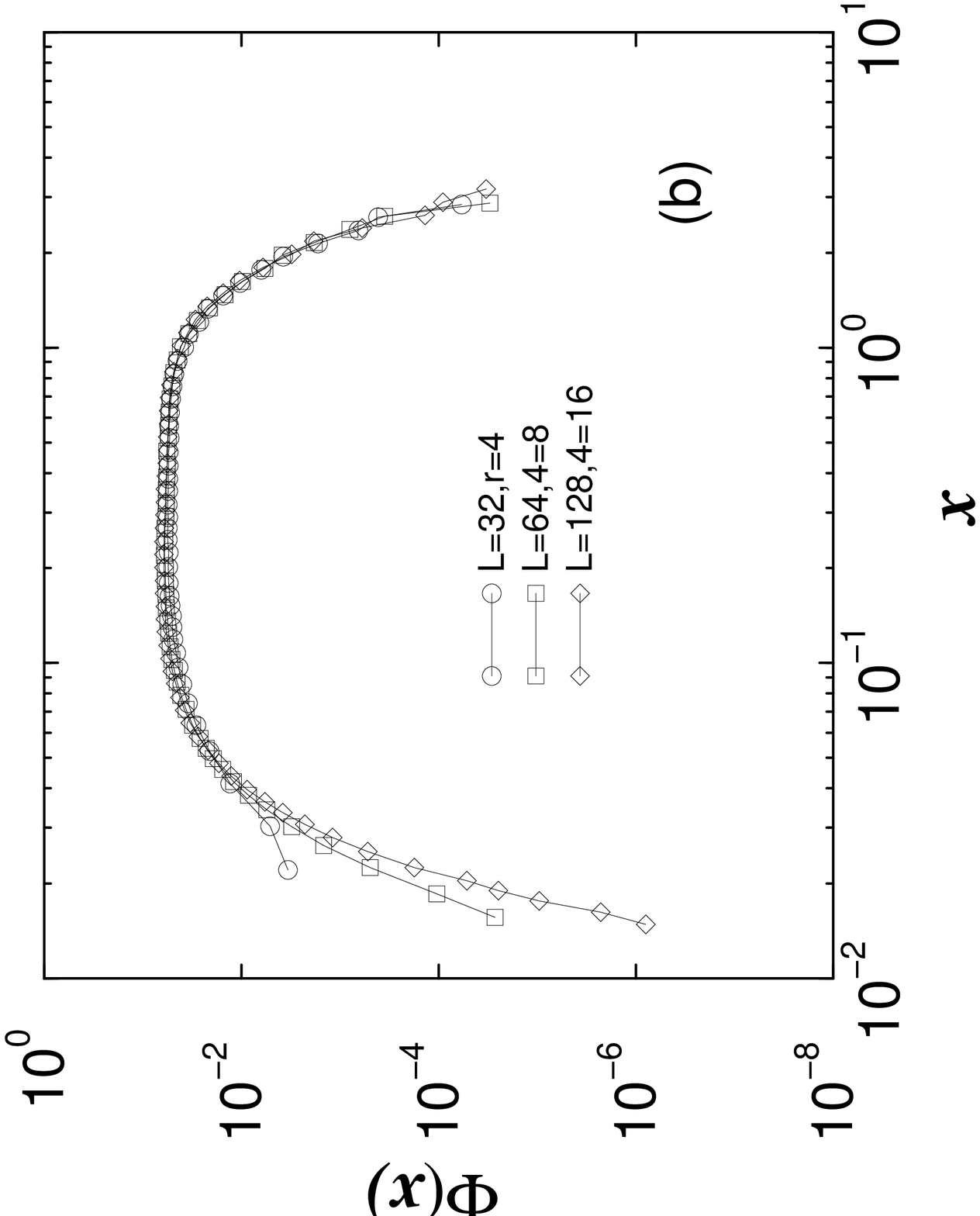}}
}
\centerline{
\epsfxsize=7.0cm
\rotate[r]{\epsfbox{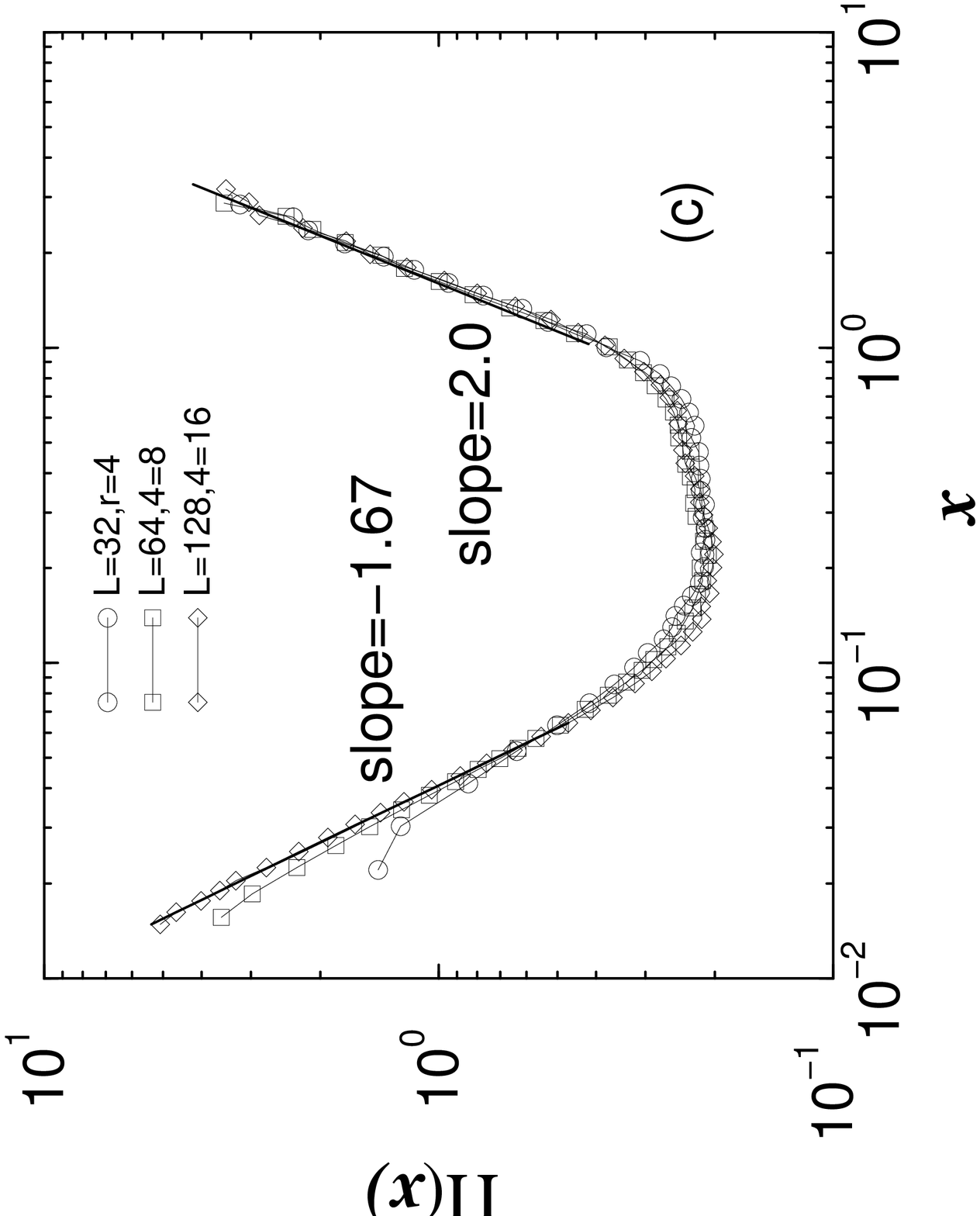}}
\vspace*{1.0cm}
}
\caption{
For $d=3$,(a) log-log plot of $P'(t|r)$ for $p=p_c=0.2488$ and for
different sets of parameters, $(r,L)=(4,32),(8,64),(16,128)$. The
power-law regime has slope $g_{t}'=2.1$. (b) Log-log plot of rescaled
probability $\Phi(x)\equiv P'(t_{\mbox{\scriptsize min}}
|r)x^{g_t'}r^{d_t}$ against rescaled length $x=t_{\mbox{\scriptsize
min}}/r^{d_t}$ using the values, $g_t'=2.1$ and $d_t=1.45$. The curves
are flat in the center because $f_2(x)$ is stretched exponential (see
Eq.~(\protect{\ref{eq:Phix}})).  (c) Log-log plot of transformed
probability $\Pi(x)=\log_{10}[A/\Phi(x)]$ versus $x=t_{\mbox{\scriptsize
min}}/r^{d_t}$. The slopes of the solid lines give the power of the
stretched exponential function $f_1$ and $f_2$ in
Eq.~(\protect{\ref{eq:Phix}}). Using the parameter $A=0.08$, the slopes
give $\phi \approx 1.6$ for the lower cut-off and $\psi \approx 2.0$ for
the upper cut-off.
}
\label{fig15}
\end{figure}

\begin{figure}[htb]
\centerline{
\epsfxsize=7.0cm
\rotate[r]{\epsfbox{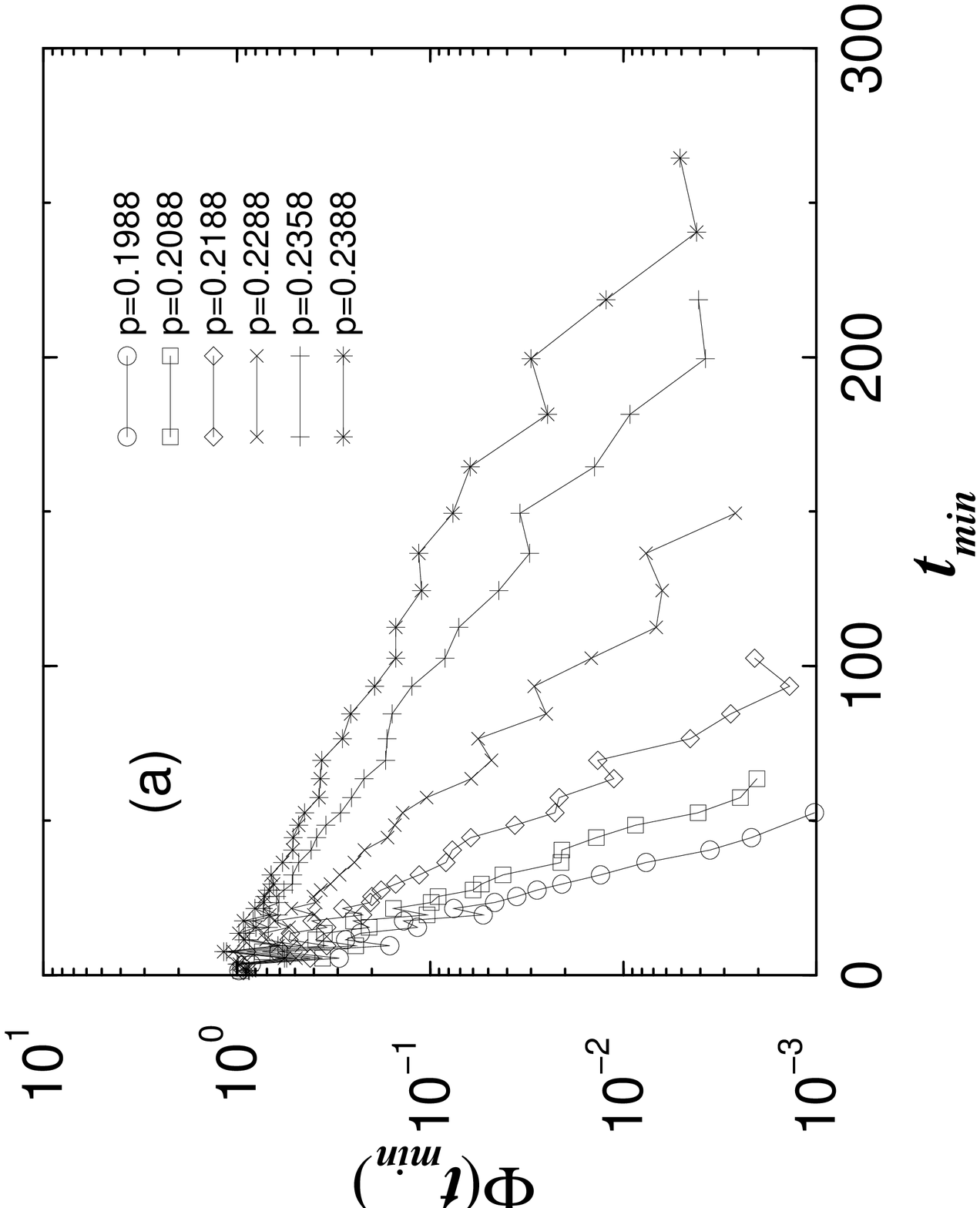}}
}
\centerline{
\epsfxsize=7.0cm
\rotate[r]{\epsfbox{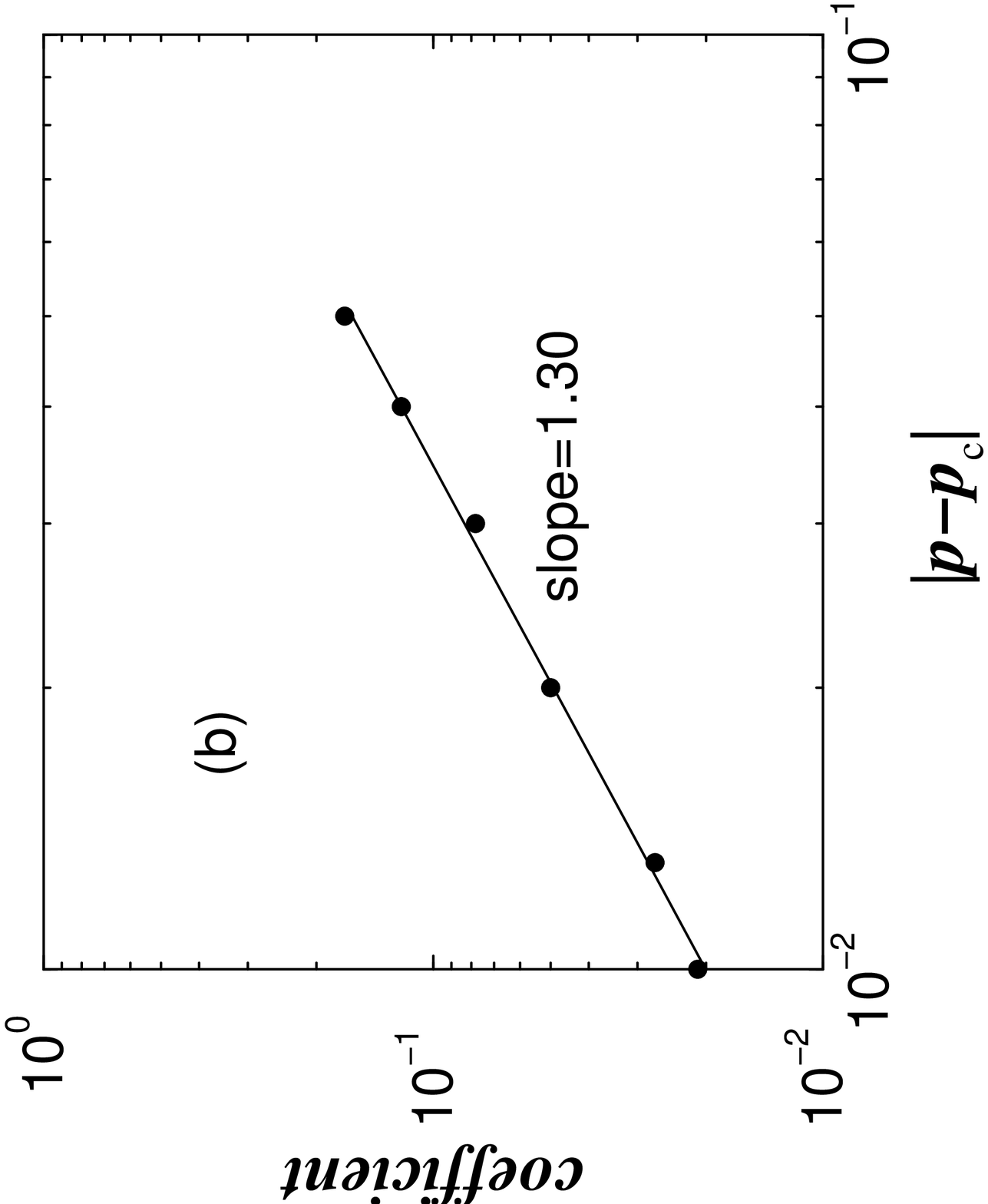}}
}
\centerline{
\epsfxsize=7.0cm
\rotate[r]{\epsfbox{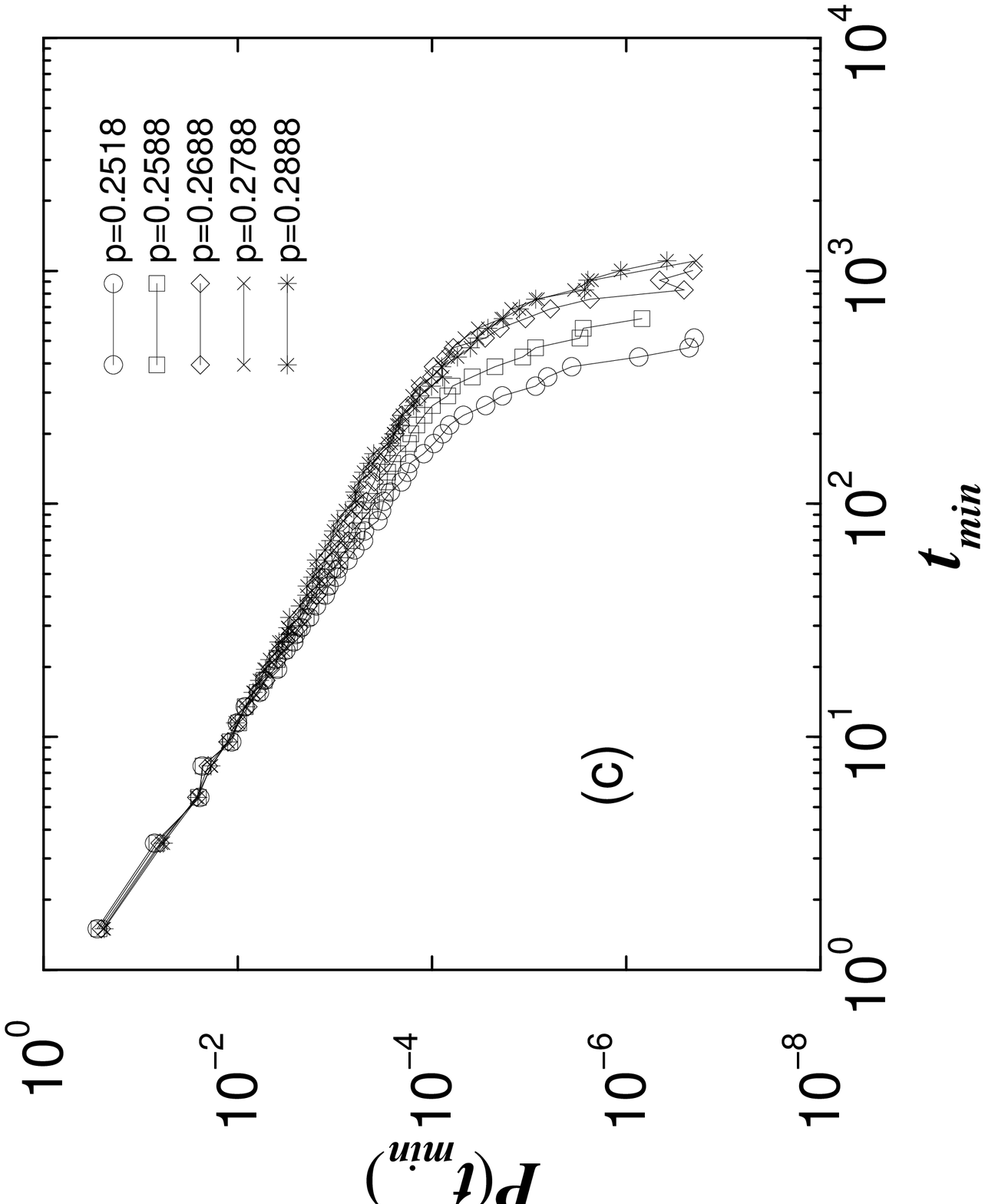}}
}
\vspace{1.0cm}
\caption{
For $d=3$, (a) semi-logarithmic plot of transformed probability
$\Phi(t_{\mbox{\scriptsize min}})$ versus $t_{\mbox{\scriptsize min}}$
below critical point for $p=0.1988,0.2088,0.2188,0.2288.0.2358,0.2388$
shows pure exponential behavior of $f_3$. (b) The slope of the log-log
plot of the coefficient in exponential function $f_3$ as a function of
$|p-p_c|$ gives the value $\nu d_t \approx 1.30$ for $p<p_c$. (c)
$P(t_{\mbox{\scriptsize min}})$ for $p>p_c$. Note that for the values of
$p$ simulated, the large $t_{\mbox{\scriptsize min}}$ behavior is
determined by the finite size of the system---not $f_3$.
}
\label{fig16}
\end{figure}

The analysis for three dimensions is completely analogous to that for
two dimensions. Our results are shown in Figs.~\ref{fig15} and
\ref{fig16} and the scaling parameters found are included in Table~I.

\subsection{Fastest Path}
\label{sec:fastestpath}

We observe that the path which takes minimal time is not always the
shortest path. However analysis of the distributions of
$\ell_{\mbox{\scriptsize min}}$ yields parameters identical to those for
the distribution of the shortest paths between points separated by
distance $r$ studied in detail in Ref.~\cite{Dokholyan98c}. Thus,
statistically, the path which takes the shortest time is one of the
paths of shortest length.

In many transport problems, the characteristic time $t^\ast$ scales with
the characteristic length $\ell^\ast$ with a power law,
\be
t^* \sim (\ell^*)^z\, .\label{eq:ts-ls}
\ee
Since $t^*$ scales as $r^{d_t}$ and $\ell^*$ scales as
$r^{d_{\mbox{\scriptsize min}}}$,
\be
z=d_t/d_{\mbox{\scriptsize min}}\, .\label{eq:z}
\ee
Since $t_{\mbox{\scriptsize min}}$ and $\ell_{\mbox{\scriptsize min}}$
are strongly correlated, the distributions $P({\ell}_{\mbox{\scriptsize min}})$ and
$P(t_{\mbox{\scriptsize min}})$ satisfy
\be
P({\ell}_{\mbox{\scriptsize min}})d{\ell_{\mbox{\scriptsize
min}}}=P(t_{\mbox{\scriptsize min}})dt_{\mbox{\scriptsize min}}
\label{eq:rel}
\ee
Combining Eqs.~(\ref{eq:ts-ls})-(\ref{eq:rel}) and the
equations for the respective distributions, we obtain a scaling
relation between exponents,
\be
(g_{\ell_{\mbox{\scriptsize min}}}-1)d_{\ell_{\mbox{\scriptsize
min}}}=(g_{\mbox{\scriptsize tm}}-1)d_{\mbox{\scriptsize
tm}}.
\label{eq:scaling}
\ee
These scaling relations are well satisfied by the set of scaling
exponents given in Table~I.

\section{Conclusions}

By modeling porous media by bond percolation and using concepts of
percolation theory, we study the flow of fluid in porous media in 2 and
3 dimensions between two ``wells'' separated by Euclidean distance
$r$. We investigate the distribution function of the shortest path
connecting the two sites, and propose a scaling {\it Ansatz\/} that
accounts for the dependence of this distribution (i) on $L$, the size of
the system, and (ii) on $p$, the bond occupancy probability. The finite
size of the system, $L$ corresponds to the size of the oil field and the
bond occupancy probability, $p$ is related to the good sand to
impermeable rock ratio of the porous media. We confirm by extensive
simulations that the {\it Ansatz\/} holds for $d=2,3$, and we calculate the
relevant scaling parameters.

In order to understand the properties of the flow of oil displaced by
fluid or gas, we study the dynamics of flow on percolation clusters. We
study two dynamical quantities: the minimal traveling time and the length of
the path corresponding to the minimal traveling time. Because of the
approximate parallel between the shortest path and the minimal traveling time
of flow, the study of the {\it shortest\/} path is the first step in
understanding the properties of the oil fields. In particular, a scaling
{\it Ansatz\/} for these dynamical quantities includes the effect of
finite system size and off-critical bond occupation probability. We find
that the scaling form for the distribution functions for these dynamical
quantities for $d=2,3$ is similar to, {\it but not identical to}, that
for the shortest path. In addition to calculating the relevant
distribution functions and scaling relations, we also calculate a number
of new exponents (see Table I).

\section{acknowledgments}

We thank BP Amoco for financial support, and M. Barth\'el\'emy,
A. Coniglio, J. Koplik, S. Redner, and R. M. Ziff for helpful
discussions.

\appendix

\section{Sampling all size clusters versus sampling only infinite
clusters}
\label{appendix_C}

Since $P'(\ell|r)$ is the distribution of shortest paths between point
in clusters of all sizes (not just the infinite cluster), we can write
\begin{equation}
\label{e1}
P'(\ell|r)=\int_0^\infty P(\ell|r,M)P(M)dM,
\end{equation}
where $P(\ell|r,M)$ is the distribution of shortest paths in clusters of
mass $M$.

The size $L$ of a cluster of mass $M$ is of the order $M^{1/d_f}$ and
the shortest path between two points in the cluster is larger than
$L^{d_{\mbox{\scriptsize min}}}$ drops off exponentially, so
$P(\ell|r,M)$ will effectively be zero when $M$ is less than
$\ell^{d_f/d_{\mbox{\scriptsize min}}}$. On the other hand, when $M$ is
greater than this value, $P(\ell|r,M)$ will be the same distribution as
the distribution for the case when the two points are in an infinite
cluster. This can be taken into account by replacing $P(\ell|r,M)$ with
$P(\ell|r)$ and replacing the zero lower limit of the integral by
$\ell^{d_f/d_{\mbox{\scriptsize min}}}$.  Since $P(M)\sim M^{-\tau}$,
where $\tau=d/d_f$, we have
\begin{eqnarray}
\label{e2}
\nonumber P'(\ell|r) &=& \int_{\ell^{d_f/d_{\mbox{\scriptsize
min}}}}^\infty P(\ell|r)M^{-d/d_f}dM \\
&=& P(\ell|r)\ell^{(d_f-d)/d_{\mbox{\scriptsize
min}}},
\end{eqnarray}
from which follows
\be
\label{e3}
g_\ell'=g_\ell+{d-d_f\over d_{\mbox{\scriptsize
min}}}=g_\ell+(d-d_f)\tilde\nu. 
\en
This can be generalized for any quantity, $x$, with fractal dimension
$d_x$ and with distributions
\begin{equation}
\label{e3x}
P(x|r)={1\over r^{d_x}}\left({x\over r^{d_x}}\right)^{-g_x}
\end{equation}
and
\begin{equation}
\label{e3y}
P'(x|r)={1\over r^{d_x}}\left({x\over r^{d_x}}\right)^{-g_x'},
\end{equation}
where ``infinite clusters'' and all finite-sized clusters, respectively,
are sampled. This generalization results in
\be
\label{e3z}
g_x'=g_x+{d-d_f\over d_x}.
\ee

\end{multicols}

\newpage

Table I. Summary of exponents and coefficients in scaling form
\[
P(x|r)\sim {(1/r^{d_x})} ({x/r^{d_x}})^{-g_x} f_1({x/r^{d_x}})
f_2({x/L^{d_x}}) f_3({x/\xi^{d_x}})
\]
where $f_1(y) = \exp(-a_xy^{-\phi_x})$, $f_2(y)=\exp(-b_xy^{\psi_x})$,
$f_3(y) = \exp(-c_xy)$.  Here $x$ denotes one of the quantities, $\ell$
or $t_{\mbox{\scriptsize min}}$. The notation N/A means Not Applicable
(since no theoretical value exists), while the notation (+/-) indicates
above or below $p_c$.

\bigskip

(a) $d=2$ results
\begin{center}
\begin{tabular}{|c||c|c||c|c||}\hline
$x$& \multicolumn{2}{c||}{$\ell$}  & \multicolumn{2}{c||}{$t_{\mbox{\scriptsize min}}$} \\ 
\cline{2-5}
exponent  & sim.            & theory           & sim.            & theory \\\hline
$d_x    $ & $1.13 \pm 0.01$ & $N/A$            & $1.33 \pm 0.05$ & $N/A$  \\\hline
$g_x'   $ & $2.14 \pm 0.02$ & 2.11             & $2.0 \pm 0.1  $ & $N/A$  \\\hline
$a_x    $ & $0.5$           & $N/A$            & $1.1$           & $N/A$  \\\hline
$\phi_x $ & $7.3\pm 0.5$    & $1/(d_x-1)=7.69$ & $3.0$           & $3.0$  \\\hline
$b_x    $ & $3.5$           & $N/A$            & $5.0$           & $N/A$  \\\hline
$\psi_x $ & $4.0\pm 0.5$    & $N/A$            & $3.0$           & $N/A$  \\\hline
$c_x    $ & $2.4(-),3.7(+)$ & $N/A$            & $1.6(-),2.6(+)$ & $N/A$  \\\hline
\end{tabular}

\vspace{0.5cm}
\end{center}

(b) $d=3$ results
\begin{center}
\begin{tabular}{|c||c|c||c|c||}\hline
$x$ & \multicolumn{2}{c||}{$\ell$}  &
\multicolumn{2}{c||}{$t_{\mbox{\scriptsize min}}$} \\ 
\cline{2-5}
exponent  & sim.            & theory           & sim.          & theory \\\hline
$d_x    $ & $1.39 \pm 0.05$ & $N/A$            &$1.45\pm0.1$   & $N/A$  \\\hline
$g_x'   $ & $2.3  \pm0.1$   & $2.23$           &$2.1\pm0.1$    & $N/A$  \\\hline
$a_x    $ & $1.4$           & $N/A$            &$2.5$          &$N/A$   \\\hline
$\phi_x $ & $2.1 \pm 0.5$   & $1/(d_x-1)=2.56$ &$1.6$          &$2.0$   \\\hline
$b_x    $ & $2.0$           & $N/A$            &$2.3$          &$N/A$   \\\hline
$\psi_x $ & $2.5 \pm 0.5$   & $N/A$            &$2.0$          &$N/A$   \\\hline
$c_x    $ & $3.1(-)$        & $N/A$            &$2.9(-)$       &$N/A$   \\\hline
\end{tabular}
\vspace{0.5cm}
\end{center}

\newpage
\newpage

\newpage

\newpage

\newpage

\newpage

\newpage

\newpage

\newpage

\newpage

\newpage

\end{document}